\documentclass[aps,prl,twocolumn,showpacs,groupedaddress,superscriptaddress,footinbib,longbibliography]{revtex4-2}

\usepackage{mathrsfs}
\usepackage{natbib,hyperref}
\setcitestyle{square,sort&compress,comma,numbers}
\hypersetup{colorlinks=true, citecolor=blue, urlcolor=blue, linkcolor=blue}
\usepackage[english]{babel}
\usepackage[utf8]{inputenc}
\usepackage{graphicx}
\usepackage[caption=false]{subfig}
\usepackage{amsmath}
\usepackage{amsfonts}
\usepackage{amssymb}
\usepackage{color,soul}
\setulcolor{red}
\usepackage{easyReview}
\usepackage{xcolor}

\DeclareMathOperator{\Tr}{tr}

\renewcommand{\vec}[1]{\mathbf{#1}}

\DeclareMathAlphabet{\Mymathbb}{U}{bbold}{m}{n}
\DeclareMathAlphabet{\mathpzc}{OT1}{pzc}{m}{it}

\newcommand{\timeSymbol}{t}
\newcommand{\Time}{\timeSymbol} 

\newcommand{\FirstSymbol}{\mathcal{G}}
\newcommand{\First}[1][]{
  \ifthenelse{\equal{#1}{}}
  {\FirstSymbol}
  {\FirstSymbol_{\scriptscriptstyle{#1}}}
}

\newcommand{\ConstGeom}[1][]{
   \ifthenelse{\equal{#1}{}}       
   {K_{\First}\,}
   {K_{\First,#1}\,}
}

\newcommand{\FirstForm}[1][]{
  \ifthenelse{\equal{#1}{}}       
  {\operatorname{I_{\point}}}             
  {\operatorname{I_{#1}}}             
}
\newcommand{\GradSymbol}{\operatorname{\mathbf{\nabla}}}

\newcommand{\GradSurf}{\GradSymbol_{\!\SurfDomain}\!}
\newcommand{\divS}{\operatorname{div}_{\! \SurfDomain}\!}

\newcommand{\GradP}{\GradSymbol_{\!\ProjMat}\!}
\newcommand{\DivP}{\operatorname{div}_{\!\ProjMat}\!}

\newcommand{\GradC}{\GradSymbol_{\!C}\!}
\newcommand{\DivC}{\operatorname{div}_{\!C}\!}

\newcommand{\GradSurfConv}[1]{\GradSymbol_{#1}}

\newcommand{\Der}{\partial}
\newcommand{\DerT}{\Der_{\Time}\,}

\newcommand{\DerTot}[2][t]
{
  \ifthenelse{\equal{#2}{}}
  {\frac{d #2}{dt}}
  {\frac{d #2}{d #1}}
}
\newcommand{\Derd}[2][]{
  \ifthenelse{\equal{#1}{}}
  {\Diffsymbol{#2}}
  {\Diffsymbol_{{#1}}{#2}}
}
\newcommand{\Diffsymbol}{\operatorname{d}\!}
\newcommand{\Diff}[2][]{
  \ifthenelse{\equal{#1}{}}
  {\Diffsymbol{#2}}
  {\Diffsymbol{#2}_{{#1}}}
}

\newcommand{\tr}{\operatorname{tr}}

\newcommand{\REALsymbol}{\mathbb{R}}
\newcommand{\REAL}[1][]{
  \ifthenelse{\equal{#1}{}}
  {\REALsymbol}
  {{\REALsymbol}^{#1}}
}
\newcommand{\EUCLsymbol}{\mathbb E}
\newcommand{\EUCL}[1][]{
  \ifthenelse{\equal{#1}{}}
  {\EUCLsymbol}
  {{\EUCLsymbol}^{#1}}
}
\newcommand{\NATURALsymbol}{\mathbb N}
\newcommand{\NATURAL}[1][]{
  \ifthenelse{\equal{#1}{}}
  {\NATURALsymbol}
  {{\NATURALsymbol}^{#1}}
}

\newcommand{\ABS}[2][]
{
  \ifthenelse{\equal{#1}{}}
  {\left| #2 \right|}
  {\left| #2 \right|_{#1}}
}

\newcommand{\NORM}[2][]
{
  \ifthenelse{\equal{#1}{}}
  {\left\| #2 \right\|}
  {\left\| #2 \right\|_{#1}}
}
\newcommand{\vertiii}[1]{{\left\vert\kern-0.25ex\left\vert\kern-0.25ex\left\vert #1 \right\vert\kern-0.25ex\right\vert\kern-0.25ex\right\vert}}
\newcommand{\BrNORM}[2][]
{
  \ifthenelse{\equal{#1}{}}
  {\vertiii{#2}}
  {\vertiii{#2}_{#1}}
}

\newcommand{\SCAL}[3][]
{
  \ifthenelse{\equal{#1}{}}
  {\left\langle{#2},{#3}\right\rangle}
  {\left\langle{#2},{#3}\right\rangle_{#1}}
}
\newcommand{\SCALF}[3][]
{
  \ifthenelse{\equal{#1}{}}
  {\left({#2},{#3}\right)}
  {\left({#2},{#3}\right)_{#1}}
}

\newcommand{\scalprodSurf}[3][]
{
  \ifthenelse{\equal{#1}{}}
  {\left\langle {#2},{#3} \right\rangle_{\scriptscriptstyle\SurfDomain}}
  {\left\langle {#2},{#3} \right\rangle_{{#1}}}
}

\newcommand{\InnerApprox}[2]{\left({#1}\,,\,{#2}\right)_{\meshparam}}

\newcommand{\point}[1][]
{
  \ifthenelse{\equal{#1}{}}
  {\mathbf{p}}
  {\mathbf{p}_{#1}}
}

\newcommand{\midPoint}{\mathbf{m}}

\newcommand{\RegionSymb}{R}
\newcommand{\Region}[1][]
{
  \ifthenelse{\equal{#1}{}}
  {\RegionSymb}
  {\RegionSymb_{#1}}
}

\newcommand{\Interval}[1][]
{
  \ifthenelse{\equal{#1}{}}
  {I}
  {I_{#1}}
}

\newcommand{\SurfDomain}{\mathcal{S}}
\newcommand{\tSurfDomain}[1][]{
  \ifthenelse{\equal{#1}{}}
  {\tilde{\SurfDomainsymb}}
  {\tilde{\SurfDomainsymb}_{#1}}
}
\newcommand{\SurfDomainBndsymb}{\partial\Gamma}
\newcommand{\SurfDomainBnd}[1][]{
  \ifthenelse{\equal{#1}{}}
  {\SurfDomainBndsymb}
  {\SurfDomainBndsymb_{#1}}
}
\newcommand{\ClosedSurfDomain}[1][]{
  \ifthenelse{\equal{#1}{}}
  {\Closedsymb{\SurfDomain}}
  {\Closedsymb{\SurfDomain[#1]}}
}

\newcommand{\heightsymb}{\mathcal{H}}
\newcommand{\height}[1][]{
  \ifthenelse{\equal{#1}{}}
  {\heightsymb}
  {\heightsymb_{#1}}
}

\newcommand{\BSMsymbol}{\mathcal{B}}

\newcommand{\BSM}[1][]
{
  \ifthenelse{\equal{#1}{}}
  {\BSMsymbol}
  {\BSMsymbol_{#1}}
}

\newcommand{\Surf}{\mathcal{S}}

\newcommand{\SurfBndsymb}{\partial\Surf}
\newcommand{\SurfBnd}[1][]{
  \ifthenelse{\equal{#1}{}}
  {\SurfBndsymb}
  {\SurfBndsymb_{#1}}
}
\newcommand{\Closedsymb}[1]{\bar{#1}}
\newcommand{\ClosedSurf}[1][]{
  \ifthenelse{\equal{#1}{}}
  {\Closedsymb{\Surf}}
  {\Closedsymb{\Surf[#1]}}
}

\newcommand{\Vector}[1]{\mathbf{#1}}

\newcommand{\press}{p}

\newcommand{\densitySymb}{\rho}
\newcommand{\density}[1][]{
  \ifthenelse{\equal{#1}{}}
  {\densitySymb}
  {\densitySymb_{\scriptscriptstyle{#1}}}
}

\newcommand{\pressApprox}{\press_{\meshparam}}

\newcommand{\bendingSymb}{\kappa}
\newcommand{\bendStiff}[1][]{
  \ifthenelse{\equal{#1}{}}
  {\bendingSymb}
  {\bendingSymb_{#1}}
}

\newcommand{\bendStiffGauss}[1][]{
  \ifthenelse{\equal{#1}{}}
  {\overline{\bendingSymb}}
  {\overline{\bendingSymb}_{#1}}
}

\newcommand{\param}{\boldsymbol{X}}

\newcommand{\update}{\boldsymbol{Y}}

\newcommand{\MapU}{\param}
\newcommand{\MapLinsymb}{F}
\newcommand{\MapLin}[1][]
{
  \ifthenelse{\equal{#1}{}}
  {\ensuremath{\MapLinsymb}}
  {\ensuremath{\MapLinsymb_{#1}}}
}

\newcommand{\MapVsymb}{\psi}
\newcommand{\MapV}[1][]
{
  \ifthenelse{\equal{#1}{}}
    {\MapVsymb}
    {\MapVsymb_{#1}}
}
\newcommand{\Transsymb}{\Phi}
\newcommand{\Trans}[1][]
{
  \ifthenelse{\equal{#1}{}}
    {\Transsymb} 
    {\Transsymb_{\scriptscriptstyle{#1}}}
}
\newcommand{\InvMapsymb}{\Psi}
\newcommand{\InvMap}[1][]
{
  \ifthenelse{\equal{#1}{}}
    {\InvMapsymb}
    {\InvMapsymb_{\scriptscriptstyle{#1}}}
}

\newcommand{\fsymb}{f}
\newcommand{\scalFun}[1][]
{
  \ifthenelse{\equal{#1}{}}
  {\fsymb}
  {\fsymb_{#1}}
}
\newcommand{\tscalFun}[1][]
{
  \ifthenelse{\equal{#1}{}}
  {\tilde{\fsymb}}
  {\tilde{\fsymb}_{#1}}
}
\newcommand{\bscalFun}[1][]
{
  \ifthenelse{\equal{#1}{}}
  {\bar{\fsymb}}
  {\bar{\fsymb}_{#1}}
}
\newcommand{\gsymb}{g}
\newcommand{\scalFung}[1][]
{
  \ifthenelse{\equal{#1}{}}
  {\gsymb}
  {\gsymb_{#1}}
}
\newcommand{\Fsymb}{F}
\newcommand{\FvecFun}[1][]
{
  \ifthenelse{\equal{#1}{}}
  {\Fsymb}
  {\Fsymb_{#1}}
}
\newcommand{\tFvecFun}[1][]
{
  \ifthenelse{\equal{#1}{}}
  {\tilde{\Fsymb}}
  {\tilde{\Fsymb}_{#1}}
}
\newcommand{\Ffunc}[2][]
{
  \ifthenelse{\equal{#1}{}}
  {\ensuremath{\Fsymb_{#2}}}
  {\ensuremath{\Fsymb^{#1}_{#2}}}
}
\newcommand{\Gsymb}{g}
\newcommand{\Gfun}[1][]
{
  \ifthenelse{\equal{#1}{}}
  {\ensuremath{\Gsymb}}
  {\ensuremath{\Gsymb_{{#1}}}}
}
\newcommand{\bGfun}[1][]
{
  \ifthenelse{\equal{#1}{}}
  {\ensuremath{\bar{\Gsymb}}}
  {\ensuremath{\bar{\Gsymb}_{{#1}}}}
}

\newcommand{\PrincipalK}[1][]
{
  \ifthenelse{\equal{#1}{}}
  {k}
  {k_{#1}}
}

\newcommand{\vecsymb}{u}
\newcommand{\vecFun}[1][]{
  \ifthenelse{\equal{#1}{}}
  {\Vector{\vecsymb}}
  {\vecsymb^{#1}}
}
\newcommand{\tvecFun}[1][]{
  \ifthenelse{\equal{#1}{}}
  {\tilde{\vecsymb}}
  {\tilde{\vecsymb}_{#1}}
}

\newcommand{\wwsymb}{w}
\newcommand{\ww}[1][]
{
  \ifthenelse{\equal{#1}{}}
  {\mathbf{\wwsymb}}
  {\wwsymb^{#1}}
}
\newcommand{\uusymb}{u}
\newcommand{\uu}[1][]
{
  \ifthenelse{\equal{#1}{}}
  {\mathbf{\uusymb}}
  {\uusymb^{#1}}
}

\newcommand{\VecFieldSymbol}{X}
\newcommand{\VecField}[1][]
{
  \ifthenelse{\equal{#1}{}}
  {\VecFieldSymbol}
  {\VecFieldSymbol^{#1}}
}
\newcommand{\VecFieldSymbolC}{Y}
\newcommand{\VecFieldYSymbol}{\boldsymbol{\VecFieldSymbolC}}
\newcommand{\VecFieldY}[1][]
{
  \ifthenelse{\equal{#1}{}}
  {\VecFieldYSymbol}
  {\VecFieldYSymbol^{#1}}
}

\newcommand{\xvsymb}{x}
\newcommand{\xv}[1][]
{
  \ifthenelse{\equal{#1}{}}
  {\mathbf{\xvsymb}}
  {\mathbf{\xvsymb}_{\scriptscriptstyle{#1}}}
}
\newcommand{\xvcomp}[1][]{
  \ifthenelse{\equal{#1}{}}
  {\xvsymb}
  {\xvsymb^{\scriptscriptstyle{#1}}}
}
\newcommand{\xcg}[1][]{
  \ifthenelse{\equal{#1}{}}
  {\xvcomp[1]}
  {\xvcomp[1]_{\scriptscriptstyle{#1}}}
}
\newcommand{\ycg}[1][]{
  \ifthenelse{\equal{#1}{}}
  {\xvcomp[2]}
  {\xvcomp[2]_{\scriptscriptstyle{#1}}}
}
\newcommand{\zcg}[1][]{
  \ifthenelse{\equal{#1}{}}
  {\xvcomp[3]}
  {\xvcomp[3]_{\scriptscriptstyle{#1}}}
}

\newcommand{\svsymb}{s}
\newcommand{\sv}[1][]{
  \ifthenelse{\equal{#1}{}}
  {\mathbf{\svsymb}}
  {\mathbf{\svsymb}_{\scriptscriptstyle{#1}}}
}
\newcommand{\svcomp}[1][]
{
  \ifthenelse{\equal{#1}{}}
  {\svsymb}
  {\svsymb^{\scriptscriptstyle{#1}}}
}
\newcommand{\xcl}[1][]{
  \ifthenelse{\equal{#1}{}}
   {\svcomp[1]}
   {\svcomp[1]_{\scriptscriptstyle{#1}}}
}
\newcommand{\ycl}[1][]{
  \ifthenelse{\equal{#1}{}}
   {\svcomp[2]}
   {\svcomp[2]_{\scriptscriptstyle{#1}}}
}
\newcommand{\zcl}[1][]{
  \ifthenelse{\equal{#1}{}}
   {\svcomp[3]}
   {\svcomp[3]_{\scriptscriptstyle{#1}}}
}

\newcommand{\ProjSymb}{\operatorname{\pi}}
\newcommand{\ProjFun}[2][]{
  \ifthenelse{\equal{#1}{}}
  {\ProjSymb\left(#2\right)}
  {\ProjSymb_{\scriptscriptstyle{#1}}\left(#2\right)}
}
\newcommand{\Prm}[1][]
{
  \ifthenelse{\equal{#1}{}}
  {\operatorname{pr}}
  {\operatorname{pr}_{\scriptscriptstyle{#1}}}
}
\newcommand{\TanPlane}[2][]
{
  \ifthenelse{\equal{#1}{}}
  {T_{\scriptscriptstyle{\point}}#2}
  {T_{\scriptscriptstyle{#1}}#2}
}

\newcommand{\SubsetSymbol}{\mathcal{U}}

\newcommand{\SubsetU}[1][]
{
  \ifthenelse{\equal{#1}{}}
  {{U}}
  {{U}_{#1}}
}
\newcommand{\SubsetV}[1][]
{
  \ifthenelse{\equal{#1}{}}
  {{V}}
  {{V}_{#1}}
}
\newcommand{\SubsetW}[1][]
{
  \ifthenelse{\equal{#1}{}}
  {{W}}
  {{W}_{#1}}
}
\newcommand{\NeighSymbol}{\mathcal{N}}
\newcommand{\Neigh}[1][]
{
  \ifthenelse{\equal{#1}{}}
  {\NeighSymbol_{\point}}
  {\NeighSymbol_{#1}}
}
\newcommand{\NeighSurf}[1][]
{
  \ifthenelse{\equal{#1}{}}
  {\SubsetSymbol_{\point}}
  {\SubsetSymbol_{#1}}
}
\newcommand{\NormSymb}{\nu}
\newcommand{\normalvec}{\boldsymbol{\nu}}
\newcommand{\normalvecApprox}{\NormSymb_{\meshparam}}
\newcommand{\normalSurf}[1][]
{
  \ifthenelse{\equal{#1}{}}
  {\NormSymb}
  {\NormSymb(#1)}
}
\newcommand{\normalInterp}[1][]
{
  \ifthenelse{\equal{#1}{}}
   {\tilde{\NormSymb}}
   {\tilde{\NormSymb}_{\scriptscriptstyle{#1}}}
}

\newcommand{\normalEdge}{\mathbf{\nu}} 
\newcommand{\basisCC}{t}
\newcommand{\basisGC}{e}
\newcommand{\vecBaseGC}[1][]
{
  \ifthenelse{\equal{#1}{}}
  {\mathbf{\basisGC}}
  {\mathbf{\basisGC}_{#1}}
}
\newcommand{\vecBasePhys}[1][]
{
  \ifthenelse{\equal{#1}{}}
  {\mathbf{\basisGC}}
  {\mathbf{\basisGC}_{#1}}
}

\newcommand{\vecBaseCCcv}[1][]
{
  \ifthenelse{\equal{#1}{}}
  {\mathbf{\basisCC}}
  {\mathbf{\basisCC}_{#1}}
}

\newcommand{\tvecBaseCCcv}[1][]
{
  \ifthenelse{\equal{#1}{}}
  {\tilde{\mathbf{\basisCC}}}
  {\tilde{\mathbf{\basisCC}}_{#1}}
}
\newcommand{\hvecBaseCCcv}[1][]
{
  \ifthenelse{\equal{#1}{}}
  {\hat{\mathbf{\basisCC}}}
  {\hat{\mathbf{\basisCC}}_{#1}}
}
\newcommand{\vecBaseCCctrv}[1][]
{
  \ifthenelse{\equal{#1}{}}
  {\mathbf{\basisCC}}
  {\mathbf{\basisCC}^{#1}}
}

\newcommand{\first}[1]{
  \IfEqCase{#1}{
    {1}{\operatorname{E}}
    {2}{\operatorname{F}}
    {3}{\operatorname{G}}
  }
  [\PackageError{first}{Undefined option to first: #1}{}]%
}
\newcommand{\SecondFormSymbol}{\ensuremath{\operatorname{II}}}
\newcommand{\SecondForm}[1][]
{
  \ifthenelse{\equal{#1}{}}
  {\SecondFormSymbol_{\point}}
  {\SecondFormSymbol_{#1}}
}
\newcommand{\second}[1]{
  \IfEqCase{#1}{
    {1}{\operatorname{e}}
    {2}{\operatorname{f}}
    {3}{\operatorname{g}}
  }
  [\PackageError{first}{Undefined option to first: #1}{}]
}
\newcommand{\WeigSymbol}{\mathcal{W}}
\newcommand{\Weig}[1][]
{
  \ifthenelse{\equal{#1}{}}
  {\WeigSymbol}
  {\WeigSymbol_{#1}}
}

\newcommand{\velSymbol}{\boldsymbol{u}}
\newcommand{\vectvel}{\mathbf{\velSymbol}}

\newcommand{\velVSymbol}{\boldsymbol{w}}
\newcommand{\vectvelV}{\mathbf{\velVSymbol}}
\newcommand{\vectvelApprox}{\mathbf{\velSymbol}_{\meshparam}}

\newcommand{\velcompContr}[2][i]
{
   \ifthenelse{\equal{#2}{}}
   {\velSymbol^{#1}}
   {\velSymbol^{#1}(#2)}}
\newcommand{\velcompPhys}[2][i]
{
   \ifthenelse{\equal{#2}{}}
   {\velSymbol_{(#1)}}
   {\velSymbol_{(#1)}(#2)}}

\newcommand{\velSymbolRP}{v}
\newcommand{\velcompContrRP}[2][i]
{
   \ifthenelse{\equal{#2}{}}
   {\velSymbolRP^{#1}}
   {\velSymbolRP^{#1}(#2)}}
\newcommand{\velRP}[1][]
{
  \ifthenelse{\equal{#1}{}}
  {\velSymbolRP}
  {\velSymbolRP_{#1}}
}

\newcommand{\velcompApprox}[2][i]
{
   \ifthenelse{\equal{#2}{}}
   {\velSymbol^{#1)}}
   {\velSymbol^{#1}_{(#2)}}
}

\newcommand{\VelSymbol}{U}
\newcommand{\vectVel}[1][]
{
   \ifthenelse{\equal{#1}{}}
   {\vec{\VelSymbol}}
   {\vec{\VelSymbol}(#1)}
}
\newcommand{\Velcomp}[2][i]
{
   \ifthenelse{\equal{#2}{}}
   {\VelSymbol^{#1}}
   {\VelSymbol^{#1}(#2)}
}
\newcommand{\VprimoSymbol}{\tilde{u}}
\newcommand{\Vprimo}[1][]
{
   \ifthenelse{\equal{#1}{}}
   {\VprimoSymbol}
   {\VprimoSymbol(#1)}
}
\newcommand{\VprimoComp}[2][i]
{
   \ifthenelse{\equal{#2}{}}
   {\VprimoSymbol^{#1}}
   {\VprimoSymbol^{#1}(#2)}
}
\newcommand{\ttvelSymbol}{\tilde{\Mymathbb{u}}}
\newcommand{\ttvel}[1][]
{
   \ifthenelse{\equal{#1}{}}
   {\mathbf{\ttvelSymbol}}
   {\mathbf{\ttvelSymbol}(#1)}
}
\newcommand{\ttvelComp}[2][i]
{
   \ifthenelse{\equal{#2}{}}
   {\ttvelSymbol^{#1}}
   {\ttvelSymbol^{#1}(#2)}
}

\newcommand{\MatAlphaSymbol}{\mathbb{A}}
\newcommand{\MatAlpha}[1][]{%
  \ifthenelse{\equal{#1}{}}
  {\MatAlphaSymbol}
  {\MatAlphaSymbol_{#1}}
}

\newcommand{\QSymbol}{q}
\newcommand{\Qdisch}[1][]
{
   \ifthenelse{\equal{#1}{}}
   {\mathbf{\QSymbol}}
   {\mathbf{\QSymbol}(#1)}
}
\newcommand{\Qcomp}[2][i]
{
   \ifthenelse{\equal{#2}{}}
   {\QSymbol^{#1}}
   {\QSymbol^{#1}(#2)}
}
\newcommand{\Qvect}[1][]
{
   \ifthenelse{\equal{#1}{}}
   {\mathbf{\QSymbol}}
   {\mathbf{\QSymbol}(#1)}
}
\newcommand{\FricSymbol}{f}
\newcommand{\vectFric}[1][]
{
   \ifthenelse{\equal{#1}{}}
   {\mathbf{\FricSymbol}}
   {\mathbf{\FricSymbol}_{\scriptscriptstyle{#1}}}
}
\newcommand{\Friccomp}[2][i]
{
   \ifthenelse{\equal{#2}{}}
   {\FricSymbol_{#1}}
   {\FricSymbol_{#1}(#2)}}
\newcommand{\BFsymbol}{\tau}
\newcommand{\BottomFriction}[1][]{
  \ifthenelse{\equal{#1}{}}
  {\BFsymbol_{b}}
  {\BFsymbol_{b}^{#1}}
}

\newcommand{\ProjMat}{\boldsymbol{P}}
\newcommand{\IDSymbol}{\boldsymbol{I}}
\newcommand{\IDtens}[1][]{
  \ifthenelse{\equal{#1}{}}
  {\IDSymbol}
  {\IDSymbol(#1)}
}
\newcommand{\IDMat}{\IDSymbol}

\newcommand{\tensSymbol}{\boldsymbol{T}}
\newcommand{\tenscompSymbol}{\tau}
\newcommand{\tens}[1][]{
  \ifthenelse{\equal{#1}{}}
  {\tensSymbol}
  {\tensSymbol(#1)}
}
\newcommand{\tenscomp}[2][ij]
{
  \ifthenelse{\equal{#2}{}}
  {\tenscompSymbol^{#1}}
  {\tenscompSymbol^{#1}(#2)}
}
\newcommand{\tensrow}[2][i]
{
  \ifthenelse{\equal{#2}{}}
  {\tensSymbol^{(#1)}}
  {\tensSymbol^{(#1)}(#2)}
}
\newcommand{\TensSymbol}{\mathbf{T}}
\newcommand{\Tens}[1][]{
  \ifthenelse{\equal{#1}{}}
  {\TensSymbol}
  {\TensSymbol_{#1}}
}

\newcommand{\TensCompSymbol}{\TensSymbol}
\newcommand{\TensComp}[2][ij]
{
  \ifthenelse{\equal{#2}{}}
  {\TensCompSymbol^{#1}}
  {\TensCompSymbol^{#1}(#2)}
}

\newcommand{\tensPrimoSymbol}{\tilde{\mathbf{\tau}}}
\newcommand{\tensPrimo}[1][]{
  \ifthenelse{\equal{#1}{}}
  {\tensPrimoSymbol}
  {\tensPrimoSymbol(#1)}
}
\newcommand{\tensPrimoCompSymbol}{\tensPrimoSymbol}
\newcommand{\tensPrimoComp}[2][ij]
{
  \ifthenelse{\equal{#2}{}}
  {\tensPrimoCompSymbol^{#1}}
  {\tensPrimoCompSymbol^{#1}(#2)}
}
\newcommand{\MCxl}[1][]{\ifthenelse{\equal{#1}{}}{h_{(1)}}{h_{(1),#1}}} 
\newcommand{\MCyl}[1][]{\ifthenelse{\equal{#1}{}}{h_{(2)}}{h_{(2),#1}}} 
\newcommand{\MCzl}[1][]{\ifthenelse{\equal{#1}{}}{h_{(3)}}{h_{(3),#1}}}

\newcommand{\MPsymb}{h}

\newcommand{\smallestH}[1][]
{
  \ifthenelse{\equal{#1}{}}
    {{l}}
    {{l}_{#1}}
}
\newcommand{\meshparam}{\MPsymb}

\newcommand{\InradiusSymbol}{r}
\newcommand{\Inradius}[1][]
{
  \ifthenelse{\equal{#1}{}}
  {\InradiusSymbol}
  {\InradiusSymbol_{\scriptscriptstyle{#1}}}
}
\newcommand{\Tsymb}{\mathcal{T}}
\newcommand{\Triang}[1][]
{
  \ifthenelse{\equal{#1}{}}
    {\Tsymb}
    {\Tsymb_{#1}}
}
\newcommand{\TriangH}[1][]
{
  \ifthenelse{\equal{#1}{}}
    {\Tsymb_{\meshparam}}
    {\Tsymb_{#1}}
}
\renewcommand{\Triang}{\TriangH}
\newcommand{\Edgesymb}{\sigma}
\newcommand{\Edge}[1][]{
  \ifthenelse{\equal{#1}{}}
    {\Edgesymb}
    {\Edgesymb_{#1}}
}
\newcommand{\EdgeH}[1][]{
  \ifthenelse{\equal{#1}{}}
    {\Edgesymb_{\meshparam}}
    {\Edgesymb_{\meshparam,#1}}
}
\newcommand{\NEdge}[1][]{
  \ifthenelse{\equal{#1}{}}
    {N_{\Edgesymb}}
    {N_{\Edgesymb({#1})}}
}
\newcommand{\Cellsymb}{T}
\newcommand{\Cell}{\Cellsymb}
\newcommand{\tCell}[1][]{
  \ifthenelse{\equal{#1}{}}
    {\tilde{\Cellsymb}}
    {\tilde{\Cellsymb}_{#1}}
}
\newcommand{\CellH}{\Cellsymb_{\meshparam}}

\newcommand{\areaSymb}{\mathcal{A}}
\newcommand{\CellArea}[1][]
{
  \ifthenelse{\equal{#1}{}}
    {\areaSymb_{\Cell}}
    {\areaSymb_{#1}}
}
\newcommand{\CellHArea}[1][]
{
  \ifthenelse{\equal{#1}{}}
    {\areaSymb_{\CellH}}
    {\areaSymb_{\meshparam,#1}}
}

\newcommand{\NCell}[1][]{
  \ifthenelse{\equal{#1}{}}
    {N_{\Cellsymb}}
    {N_{\Cellsymb({#1})}}
}

\newcommand{\lengthSymb}{l}

\newcommand{\edgeLength}[1][]
{
  \ifthenelse{\equal{#1}{}}
  {\lengthSymb_{\Edge}}
  {\lengthSymb_{#1}}
}
 
\newcommand{\edgeHLength}[1][]
{
  \ifthenelse{\equal{#1}{}}
  {\lengthSymb_{\EdgeH}}
  {\lengthSymb_{\meshparam,#1}}
}

\newcommand{\Sourcesymb}{\mathbf{S}}
\newcommand{\Source}[1][]
{
  \ifthenelse{\equal{#1}{}}
    {\Sourcesymb}
    {\Sourcesymb_{#1}}
}

\newcommand{\SourceEdge}[1][]{
  \ifthenelse{\equal{#1}{}}
    {\Sourcesymb_{ij}}
    {\Sourcesymb_{#1}}
}

\newcommand{\FluxEdgesymb}{\mathbf{F}}
\newcommand{\FluxEdge}[1][]{
  \ifthenelse{\equal{#1}{}}
    {\FluxEdgesymb_{ij}}
    {\FluxEdgesymb_{#1}}
}

\newcommand{\numFlux}[1][]
{
  \ifthenelse{\equal{#1}{}}
  {\tilde{\fsymb}}
  {\tilde{\fsymb}_{#1}}
}

\newcommand{\FluxFuncNormSymbol}{\mathbf{F}}
\newcommand{\FluxFuncNorm}[1][]{
  \ifthenelse{\equal{#1}{}}
    {\FluxFuncNormSymbol^{\normalEdge}}
    {\FluxFuncNormSymbol^{\normalEdge}_{#1}}    
}

\newcommand{\JacobianSymbol}{\mathbf{A}}
\newcommand{\Jacobian}[1][]{
  \ifthenelse{\equal{#1}{}}
    {\JacobianSymbol}
    {\JacobianSymbol_{#1}}    
}
\newcommand{\EValSymbol}{\lambda}
\newcommand{\EVal}[1][]{
  \ifthenelse{\equal{#1}{}}
  {\EValSymbol}
  {\EValSymbol_{#1}}
}
\newcommand{\EVecSymbol}{\mathbf{r}}
\newcommand{\EVec}[2][]{
  \ifthenelse{\equal{#1}{}}
  {\EVecSymbol^{(#2)}}
  {\EVecSymbol^{(#2)}_{#1}}
}

\newcommand{\midPointEdge}[1][]
{
  \ifthenelse{\equal{#1}{}}
  {\midPoint_{\scriptscriptstyle\Edge}}
  {\midPoint_{\scriptscriptstyle\Edge[#1]}}
}
\newcommand{\gpPointEdgeDG}[1][]
{
  \ifthenelse{\equal{#1}{}}
  {\point_{\scriptscriptstyle\Edge}}
  {\point_{\scriptscriptstyle\Edge,#1}}
}
\newcommand{\midPointCell}[1][]
{
  \ifthenelse{\equal{#1}{}}
  {\midPoint_{\scriptscriptstyle\Cell}}
  {\midPoint_{\scriptscriptstyle\Cell[#1]}}
}

\newcommand{\energy}{\mathcal{F}}

\newcommand{\speedRS}[2][]
{
  \ifthenelse{\equal{#2}{}}
  {S_{#2}}
  {S_{#2}^{#1}}
}

\newcommand{\ContSymbol}{C}
\newcommand{\Cont}[1][]{
  \ifthenelse{\equal{#1}{}}
  {\ContSymbol^{0}}
  {\ContSymbol^{#1}}
}
\newcommand{\Cinf}[1][]{
  \ifthenelse{\equal{#1}{}}
  {\ContSymbol^{\infty}}
  {\ContSymbol^{\infty}(#1)}
}
\newcommand{\SobSymbol}{W}
\newcommand{\HilbSymbol}{H}
\newcommand{\Hilb}[1][]{
  \ifthenelse{\equal{#1}{}}
  {\HilbSymbol^{1}}
  {\SobSymbol^{1}_{#1}}  
}
\newcommand{\Sob}[2][]{
  \ifthenelse{\equal{#1}{}}
  {\HilbSymbol^{#2}}
  {\SobSymbol^{#2,#1}}  
}

\newcommand{\LspaceSymb}{L}
\newcommand{\Lspace}[1][]{
  \ifthenelse{\equal{#1}{}}
  {\LspaceSymb^{2}}
  {\LspaceSymb^{#1}}  
}

\newcommand{\TestSpSymbol}{{V}}
\newcommand{\TestSpace}[1][]{
  \ifthenelse{\equal{#1}{}}
  {\TestSpSymbol({\SurfDomain})}
  {\TestSpSymbol_{#1}({\SurfDomain})}
}

\newcommand{\TestSpaceEmbedded}[1][]{
  \ifthenelse{\equal{#1}{}}
  {\TestSpSymbol(\TriangH(\SurfDomain))}
  {\TestSpSymbol_{#1}(\TriangH(\SurfDomain))}
}
\newcommand{\TestSpaceIntrinsic}[1][]{
  \ifthenelse{\equal{#1}{}}
  {\TestSpSymbol(\Triang(\SurfDomain))}
  {\TestSpSymbol_{#1}(\Triang(\SurfDomain))}
}
\newcommand{\TestSpaceChart}[1][]{
  \ifthenelse{\equal{#1}{}}
  {\TestSpSymbol(\Triang(\SubsetU))}
  {\TestSpSymbol_{#1}(\Triang(\SubsetU))}
}
\newcommand{\TestSpaceCell}[1][]{
  \ifthenelse{\equal{#1}{}}
  {\TestSpSymbol_{\meshparam}(\Cell)}
  {\TestSpSymbol_{\meshparam}(\Cell_{#1})}
}

\newcommand{\TestSpaceApprox}[1][]{
  \ifthenelse{\equal{#1}{}}
  {\TestSpSymbol_{\meshparam}}
  {\TestSpSymbol_{\meshparam}(#1)}
}
\newcommand{\TestSpaceVec}[1][]{
  \ifthenelse{\equal{#1}{}}
  {\mathbf{\TestSpSymbol}_{\meshparam}}
  {\mathbf{\TestSpSymbol}_{\meshparam}(#1)}
}

\newcommand{\TestSpGammaSymbol}{{P}}
\newcommand{\TestSpaceGamma}[1][]{
  \ifthenelse{\equal{#1}{}}
  {\TestSpGammaSymbol_{\meshparam}}
  {\TestSpGammaSymbol_{\meshparam}(#1)}
}

\newcommand{\TestSpCHSymbol}{{W}}
\newcommand{\TestSpaceCH}[1][]{
  \ifthenelse{\equal{#1}{}}
  {\TestSpCHSymbol_{\meshparam}}
  {\TestSpCHSymbol_{\meshparam}(#1)}
}

\newcommand{\TestSpPressSymbol}{Q}
\newcommand{\TestSpacePress}[1][]{
  \ifthenelse{\equal{#1}{}}
  {\TestSpPressSymbol_{\meshparam}}
  {\TestSpPressSymbol_{\meshparam}(#1)}
}

\newcommand{\GaussCurv}{\mathcal{K}}
\newcommand{\calH}{\mathcal{H}}
\newcommand{\meanCurv}{\calH}

\newcommand{\hmeanCurv}[1][]
{
  \ifthenelse{\equal{#1}{}}
  {\hat{\calH}}
  {\hat{\calH}_{#1}}
}

\newcommand{\shapeOp}{\mathcal{B}}
\newcommand{\AngleSymbol}{\theta}
\newcommand{\DevAngle}[1][]
{
  \ifthenelse{\equal{#1}{}}
  {\AngleSymbol}
  {\AngleSymbol_{\scriptscriptstyle{#1}}}
}
\newcommand{\relheightSymb}{\pi}
\newcommand{\relheight}[1][]
{
  \ifthenelse{\equal{#1}{}}
  {\relheightSymb_{\scriptscriptstyle{\SurfDomain}}}
  {\relheightSymb_{\scriptscriptstyle{#1}}}
}

\newcommand{\StressTensComp}{\sigma}
\newcommand{\StressTens}{\boldsymbol{\StressTensComp}}

\newcommand{\DiffEig}[1][]
{ \ifthenelse{\equal{#1}{}}
  {d}
  {d_{#1}}
}

\newcommand{\BilinearStiffSymbol}{a}
\newcommand{\BilinearStiff}[3][]
{
  \ifthenelse{\equal{#1}{}}
  {\BilinearStiffSymbol(#2,#3)}    
  {\BilinearStiffSymbol_{#1}(#2,#3)}    
}
\newcommand{\BilinearAdvSymbol}{b}
\newcommand{\BilinearAdv}[3][]
{
  \ifthenelse{\equal{#1}{}}
  {\BilinearAdvSymbol(#2,#3)}    
  {\BilinearAdvSymbol_{#1}(#2,#3)}    
}
\newcommand{\BilinearMassSymbol}{m}
\newcommand{\BilinearMass}[3][]
{
  \ifthenelse{\equal{#1}{}}
  {\BilinearMassSymbol(#2,#3)}    
  {\BilinearMassSymbol_{#1}(#2,#3)}    
}
\newcommand{\BilinearReactSymbol}{c}
\newcommand{\BilinearReact}[3][]
{
  \ifthenelse{\equal{#1}{}}
  {\BilinearReactSymbol(#2,#3)}    
  {\BilinearReactSymbol_{#1}(#2,#3)}    
}

\newcommand{\TestSymbol}{v}
\newcommand{\Test}[1][]
{
  \ifthenelse{\equal{#1}{}}
  {\TestSymbol}  
  {\TestSymbol_{\scriptscriptstyle{#1}}}
}

\newcommand{\nNodes}[1][]
{
  \ifthenelse{\equal{#1}{}}
  {N^{\scriptscriptstyle{dof}}}
  {N^{\scriptscriptstyle{dof}}_{\scriptscriptstyle{#1}}}
}

\newcommand{\TestApprox}[1][]
{
  \ifthenelse{\equal{#1}{}}
  {\TestSymbol_{\scriptscriptstyle{\meshparam}}}  
  {\TestSymbol_{\scriptscriptstyle{\meshparam,#1}}}
}

\newcommand{\bTestApprox}[1][]
{
  \ifthenelse{\equal{#1}{}}
  {\bar{\TestSymbol}_{\scriptscriptstyle{\meshparam}}}  
  {\bar{\TestSymbol}_{\scriptscriptstyle{\meshparam,#1}}}
}

\newcommand{\TestZSymbol}{\mathbf{Z}}
\newcommand{\TestZ}[1][]
{
  \ifthenelse{\equal{#1}{}}
  {\TestZSymbol}  
  {\TestZSymbol_{\scriptscriptstyle{#1}}}
}

\newcommand{\TestZApprox}{\TestZSymbol_{\scriptscriptstyle{\meshparam}}} 

\newcommand{\TestHSymbol}{{h}}
\newcommand{\TestH}[1][]
{
  \ifthenelse{\equal{#1}{}}
  {\TestHSymbol}  
  {\TestHSymbol_{\scriptscriptstyle{#1}}}
}

\newcommand{\TestHApprox}{\TestHSymbol_{\scriptscriptstyle{\meshparam}}}
\newcommand{\TestWSymbol}{w}
\newcommand{\TestW}[1][]
{
  \ifthenelse{\equal{#1}{}}
  {\TestWSymbol}  
  {\TestWSymbol_{\scriptscriptstyle{#1}}}
}

\newcommand{\TestWApprox}[1][]
{
  \ifthenelse{\equal{#1}{}}
  {\TestWSymbol_{\scriptscriptstyle{\meshparam}}}  
  {\TestWSymbol_{\scriptscriptstyle{\meshparam,#1}}}
}
\newcommand{\bTestWApprox}[1][]
{
  \ifthenelse{\equal{#1}{}}
  {\bar{\TestWSymbol}_{\scriptscriptstyle{\meshparam}}}  
  {\bar{\TestWSymbol}_{\scriptscriptstyle{\meshparam,#1}}}
}

\newcommand{\TestCHSymbol}{\psi}
\newcommand{\TestCHApprox}[1][]
{
  \ifthenelse{\equal{#1}{}}
  {\TestCHSymbol_{\scriptscriptstyle{\meshparam}}}  
  {\TestCHSymbol_{\scriptscriptstyle{\meshparam,#1}}}
}
\newcommand{\TestCH}[1][]
{
  \ifthenelse{\equal{#1}{}}
  {\TestCHSymbol}  
  {\TestCHSymbol_{\scriptscriptstyle{#1}}}
}
\newcommand{\TestCHmuSymbol}{\xi}
\newcommand{\TestCHmuApprox}[1][]
{
  \ifthenelse{\equal{#1}{}}
  {\TestCHmuSymbol_{\scriptscriptstyle{\meshparam}}}  
  {\TestCHmuSymbol_{\scriptscriptstyle{\meshparam,#1}}}
}

\newcommand{\TestVecSymbol}{\boldsymbol{v}}
\newcommand{\TestVecApprox}{\TestVecSymbol_{\scriptscriptstyle{\meshparam}}}
\newcommand{\TestPressSymbol}{q}
\newcommand{\TestPress}[1][]
{
  \ifthenelse{\equal{#1}{}}
  {\TestPressSymbol}  
  {\TestPressSymbol_{\scriptscriptstyle{#1}}}
}
\newcommand{\TestPressApprox} {\TestPressSymbol_{\scriptscriptstyle{\meshparam}}}

\newcommand{\viscositySymb}{\eta}
\newcommand{\viscosity}[1][]
{
  \ifthenelse{\equal{#1}{}}
  {\viscositySymb}
  {\viscositySymb_{\scriptscriptstyle{#1}}}
}

\newcommand{\ResidualSymbol}{R}
\newcommand{\Residual}[1][]
{
  \ifthenelse{\equal{#1}{}}
  {\ResidualSymbol}
  {\ResidualSymbol_{#1}}
}

\newcommand{\curvature}[1][]
{
  \ifthenelse{\equal{#1}{}}
  {\kappa}
  {\kappa_{#1}}
}

\newcommand{\QuadRule}[2][]
{
  \ifthenelse{\equal{#1}{}}
  {Q(#2)}
  {Q_{#1}(#2)}
}

\newcommand{\wrt}{w.\,r.\,t.}

\newcommand{\formComma}{\,\text{,}}
\newcommand{\formPeriod}{\,\text{.}}

\newcommand{\innerLTwo}[2]{\left\langle #2 \right\rangle_{L^2(#1)}}
\newcommand{\tangentR}[1][]{T^{#1}\R^3\vert_{\SurfDomain}}
\newcommand{\Wb}{\bm{W}}
\newcommand{\nub}{\bm{\nu}}
\newcommand{\dbdot}{\operatorname{:}}
 
\renewcommand{\S}{\mathcal{S}}
\renewcommand{\H}{\mathcal{H}}
\newcommand{\K}{\mathcal{K}}
\newcommand{\B}{\mathcal{B}}

\renewcommand{\div}{\text{div}}
\newcommand{\Reyn}{\text{Re}}

\newcommand{\densityBE}{f_{BE}}
\newcommand{\N}{\mathbb{N}}
\newcommand{\R}{\mathbb{R}}

\begin{document}

\title{The influence of higher order geometric terms on the asymmetry and dynamics of membranes}

\author{Jan Magnus Sischka}
\affiliation{Technische Universit\"at Dresden, 01062 Dresden, Germany}
\author{Ingo Nitschke}
\affiliation{Technische Universit\"at Dresden, 01062 Dresden, Germany}
\author{Axel Voigt}
\affiliation{Technische Universit\"at Dresden, 01062 Dresden, Germany}
\begin{abstract}
We consider membranes as fluid deformable surface and allow for higher order geometric terms in the bending energy. The evolution equations are derived and numerically solved using surface finite elements. The higher order geometric terms related to the Gaussian curvature squared have a tendency to stabilize tubes and enhance the evolution towards equilibrium shapes, thereby facilitating rapid shape changes. This is demonstrated in axisymmetric settings and fully three-dimensional simulations.
\end{abstract} 
\maketitle

\section{Introduction}

Membranes are ubiquitous and essential in biology, they compartmentalize biomaterials, separate the cell from its
exterior and organelles from the cytoplasm, dynamically remodel and change conformation. Geometric properties of the membrane have been identified as key players for such processes \cite{mcmahon2005membrane,Deserno_2015,D0SM90234A}. As the typical thickness of a membrane is orders of magnitude smaller than its lateral extension, treating the membrane as a two-dimensional surface embedded in a three-dimensional space is a reasonable approximation. This separation of length scales allows for a mesoscopic modeling where details related to membrane molecular structure are considered in effective material parameters and geometric quantities and led to the success of the classical Canham-Helfrich model \cite{Helf73,Canh70}, which builds on a bending energy 
$
\energy_{BE}(\meanCurv,  \K ) = \int_\mathcal{\SurfDomain} k_{2,0}(\meanCurv-\meanCurv_{0})^2 + k_{2,1}\K\,d\SurfDomain
$
with mean curvature $\H$, Gaussian curvature $\K$, bending rigidity parameters $k_{2,0}$ and $k_{2,1}$, a spontaneous curvature $\H_0$ and additional (local or global) area and volume constraints. For definitions see Section \ref{ss:notation}. Assuming constant values of $k_{2,1}$, the second term reduces to a topological measure and thus a constant as long as the topology does not change. We therefore neglect this term.

Equilibrium shapes, resulting from minimizing the bending energy $\energy_{BE}$, have been extensively studied, see \cite{Seifert_AP_1997} for a review. For $\H_0 = 0$ and wide ranges of the reduced volume, which is the ratio of the volume and the volume of an equivalent sphere with the same area, they are dominated by prolate and oblate shapes \cite{Seifert1991}. The spontaneous curvature $\H_0$, which accounts for the asymmetry of the membrane is able to modify these shapes \cite{Seifert1991}. 

A wealth of studies aims to produce tubes as minimizing shapes. Tubes are ubiquitous in membranes and play crucial roles in trafficking, ion transport, and cellular motility. For idealized situations this is rather simple as $\energy_{BE} = 0$ if $\H = \H_0$, which is achieved for an infinite tube of radius $r$ and $\H_0 = 1/r$. However, this solution is not unique, a sphere of radius $2 r$ also leads to $\energy_{BE} = 0$. More realistic situations with finite volume and area require further considerations, e.g., introducing a spontaneous curvature deviator \cite{bobrovska2013role,PhysRevE.89.062715,mahapatra2023formation}. 
Also various ideas have been proposed to consider higher order geometric terms in the bending energy $\energy_{BE}$ to enforce the stability of tubes \cite{Mitov_1978,Fournier_1997,Capovilla_2003,Shemesh_2003,Deserno_2015}. E.g., forth order terms proportional to $\K^2$ seem plausible as $\K = 0$ for tubes. 

Also the formation of tubes is considered extensively. Tube formation is attributed to the interaction of membranes with proteins that
induce curvature \cite{SIMUNOVIC2015780} or to forces exerted on the membrane \cite{PhysRevLett.88.238101,roux2002minimal}, among other
mechanisms. Within the context of the Canham-Helfrich model the first approach locally modifies $\H_0$ depending on the spatial distribution of proteins leading to so-called Jülicher-Lipowski models \cite{PhysRevE.53.2670} or variations of it \cite{PhysRevLett.99.088101,PhysRevE.79.031926,HMLLRV_IJBB_2013,doi:10.1098/rstb.2017.0115,elliott2016variational,Bachini_2023}. The second considers the relationship between the applied force resulting from the cellular cortex or external forces and the membrane properties. Of special interest are situations where tubes form spontaneously. Assuming $\H \neq \H_0$ for some part of the surface $\S_0$, one might expand the bending energy $\energy_{BE}$ on this part as $\int_{\mathcal{\SurfDomain}_0} k_{2,0} \H^2 \,d\SurfDomain - \int_{\mathcal{\SurfDomain}_0} 2 k_{2,0} \H_0 \H \,d\SurfDomain + \int_{\mathcal{\SurfDomain}_0} k_{2,0} \H_0^2 \,d\SurfDomain$. In situations for which the first two terms approximately cancel, the bending energy is just $ \int_{\mathcal{\SurfDomain}_0} k_{2,0} \H_0^2 \,d\SurfDomain$ and can thus be considered as resulting from an effective (spontaneous) surface tension $k_{2,0} \H_0^2$, which is identified as the main driving force for the spontaneous tubulation of membranes \cite{C2FD20105D}.

However, most of these studies only focus on equilibrium shapes, comparing the bending energy $\energy_{BE}$ of different configurations and addressing their stability. The dynamic evolution of the membrane is less considered. This is surprising, as also the dynamics of membranes is fascinating. Cells rapidly change shape and according to \cite{de2024follow} the membrane and the underlying cortex act as an integrated system to globally coordinate changes in cell shape. To facilitate these rapid morphological changes, cells maintain an excess of membrane that is organized in membrane reservoirs and is available to the cell in the order of seconds \cite{DeBelly_Cell_2023}. To understand such processes thus not only requires to unveil the secretes of spontaneous tubulation but also to consider the flow of membranes which facilitates the rapid shape changes. 

Surface viscosity has been identified as a key player \cite{FAIZI2022910} and considering membranes as fluid deformable surfaces \cite{ArroyoDesimone2009,Torres-Sanchez_2019,voigt2019fluid,al2021active} opened new perspectives on the description of the dynamics of membranes. Fluid deformable surfaces can be viewed as two-dimensional viscous fluids with bending elasticity. Due to this solid-fluid duality, any shape change contributes to tangential flow, and vice versa, any tangential flow on a curved surface induces shape deformations. This tight coupling between shape and flow makes curvature a natural element of the governing equations. As demonstrated by numerical studies of the equations for fluid deformable surfaces, surface hydrodynamics can significantly speed up the evolution \cite{Reuther_2020,Krause_2023} and can enhance bulging and furrow formation in membranes \cite{Bachini_2023}. 

While these models certainly cannot fully describe the formation of membrane reservoirs or their release, as the interaction with the cellular cortex is not considered, these membrane properties facilitate such processes \cite{de2024follow}. Even if various models for the cellular cortex exist \cite{10.7554/eLife.04165,reymann2016cortical,da2022viscous,bhatnagar2023axis}, we here refrain from such couplings and aim for a minimal model of the membrane which accounts for surface hydrodynamics and higher order geometric terms in the bending energy. We computationally explore the effect of these terms on the equilibrium shapes and the dynamics to reach them.

The remaining of the paper is structured as follows. In Section \ref{s:model} we introduce the full model and briefly mention the considered numerical approach. More details on the derivation of the model and on the numerical approach including convergence studies are provided in Appendices \ref{app:derivation}, \ref{app:numerics} and \ref{app:validation}. Results are described in Section \ref{s:results}. They contain axisymmetric and full three-dimensional computations addressing the dynamic evolution and the equilibrium shapes. In Section \ref{s:conclusion} we draw conclusions.

\section{Model} \label{s:model}
\subsection{Notation} \label{ss:notation}
We follow the same notation as in \cite{Bachini_2023}, which is here repeated for convenience. We consider a time dependent smooth and oriented surface $\SurfDomain = \SurfDomain(t)$ without boundary, embedded in $\R^3$. The enclosed volume is denoted by $\Omega = \Omega(t)$. We denote by $\normalvec$ the outward pointing surface normal, the surface projection is $\ProjMat=\IDMat-\normalvec\otimes\normalvec$, with $\IDMat$ the identity matrix, the shape operator is $\shapeOp= -\GradP\normalvec$, the mean curvature $\meanCurv= \text{tr}\shapeOp$, and the Gaussian curvature $\K = \frac{1}{2}\left(\H^2-\|\B\|^2\right)$. We consider time-dependent Euclidean-based $ n $-tensor fields in $T^n\R^3\vert_{\SurfDomain}$. We call $T^0\R^3\vert_{\SurfDomain} = T^0\SurfDomain$ the space of scalar fields, $T^1\R^3\vert_{\SurfDomain}=T\R^3\vert_{\SurfDomain}$ the space of vector fields, and $T^2\R^3\vert_{\SurfDomain} $ the space of 2-tensor fields.
Important subtensor fields are tangential n-tensor fields in $T^n\SurfDomain \le T^n\R^3\vert_{\SurfDomain}$. 
Let $p \in T^0\SurfDomain$ be a continuously differentiable scalar field, $\vectvel \in T\R^3\vert_{\SurfDomain}$ a continuously differentiable $\R^3$-vector field, and $\StressTens \in T^2\R^3\vert_{\SurfDomain}$ a continuously differentiable $\R^{3\times3}$-tensor field defined on $\SurfDomain$. We define the different surface gradients by $\GradP p = \ProjMat\nabla p^e$, $\GradP\vectvel = \ProjMat\nabla\vectvel^e\ProjMat$ and $\GradC\StressTens = \nabla\StressTens^e \ProjMat$,
where $p^e$, $\vectvel^e$ and $\StressTens^e$ are arbitrary smooth extensions of $p$, $\vectvel$ and $\StressTens$ in the normal direction and $\nabla$ is the gradient of the embedding space $\R^3$. The corresponding divergence operators for a vector field $\vectvel$ and a tensor field $\StressTens$ are $\DivP\vectvel = \operatorname{tr}(\GradP\vectvel)$ and $\DivC(\StressTens\ProjMat) = \operatorname{tr}\GradC(\StressTens\ProjMat)$, where $\operatorname{tr}$ is the trace operator. The relations to the covariant derivatives $\GradSurf$ and the covariant divergence $\divS$ on $\SurfDomain$, with $\Delta_{\S} = \divS \cdot \GradSurf$ the Laplace-Beltrami operator, read $\GradP p=\GradSurf p$ and
${\DivP\vectvel = \divS(\ProjMat\vectvel)-(\vectvel\cdot\normalvec)\meanCurv}$, respectively.

\subsection{Governing equations}
The material velocity $\vectvel \in T\R^3\vert_{\SurfDomain}$ can be decomposed into $\vectvel = u_N \normalvec + \vectvel_T$, with $u_N = \vectvel \cdot \normalvec$ and $\vectvel_T = \ProjMat \vectvel$, the normal and the tangential part, respectively. The pressure $p \in T^0 \SurfDomain$ serves as Lagrange multiplier for the inextensibility constraint. The governing equations for these unknowns read
\begin{align} \label{eq:dyn_sys1}
  \partial_t\mathbf{u}+\nabla_\mathbf{w}\mathbf{u} &=
   -\nabla_\S p-p\H\bm{\nu} + \frac{2}{\Reyn}\div_C\bm{\sigma} \nonumber \\
   & \quad - \gamma \bm{u} + \bm{b} - \lambda\bm{\nu}\\
   \div_{\bm{P}} \bm{u} &= 0 \\
   \label{eq:dyn_sys3}
   \int_\S \bm{u}\cdot\bm{\nu}\,d\S &= 0 \,,
\end{align}
where $[\nabla_\mathbf{w}\vectvel]_i = (\GradSurf \vectvel_i, \mathbf{w})$, $i = 1,2,3$, with $\mathbf{w} = \vectvel - \partial_t \mathbf{X}$ is the relative velocity and $\mathbf{X}$ a parameterisation of $\SurfDomain$, $\StressTens(\vectvel) = \frac{1}{2} (\GradP \vectvel + (\GradP \vectvel)^T) \in T^2\R^3\vert_{\SurfDomain}$ is the rate of deformation tensor, $\Reyn>0$ is the Reynolds number, $\gamma\geq0$ is a friction coefficient, $\lambda \in \R$ is a Lagrange multiplier to ensure a constant enclosed volume, and $\mathbf{b}$ denotes a bending force, defined by
\begin{align} 
  \bm{b} =& -2k_{2,0}(\Delta_\S\meanCurv+(\meanCurv-\meanCurv_0)(\|\shapeOp\|^2-\frac{1}{2}\meanCurv(\meanCurv-\meanCurv_0)))\normalvec\nonumber\\
         & - 12k_{4,0} \div_\S ( (\meanCurv - \meanCurv_{0})^2 \nabla_{\SurfDomain}\meanCurv )\normalvec	\nonumber\\
         & - k_{4,0} ((\meanCurv - \meanCurv_{0}3\meanCurv+\meanCurv_{0} )\meanCurv - 8\K ) )\normalvec\nonumber\\
         &-k_{4,2}(2\div_\S ( ( \meanCurv\ProjMat-\shapeOp )\nabla_{\SurfDomain}\K ) + \meanCurv \K^2 )\normalvec,\label{eq:bending_force}
\end{align}
where $\meanCurv_0$ is a spontaneous curvature and $k_{2,0}, k_{4,0},  k_{4,2}\in\R$ are bending rigidity parameters. The system of equations consider a model for fluid deformable surfaces. Such models consist of incompressible surface Navier-Stokes equations with
bending forces and a constraint on the enclosed volume. The highly nonlinear model accounts for the tight interplay between surface evolution, surface curvature and surface hydrodynamics and allows to model membranes with surface viscosity. For derivations of the model (with $k_{4,0} = k_{4,2} = 0$) we refer to \cite{Torres-Sanchez_2019,nitschke2019hydrodynamic,Reuther_2020,Bachini_2023,NitschkeVoigt_2025}. They consider different principle and build on a nonlinear Onsager formalism \cite{Torres-Sanchez_2019}, thin film limits from three-dimensional models \cite{nitschke2019hydrodynamic,Reuther_2020} and a Lagrange-D'Alembert approach \cite{Bachini_2023,NitschkeVoigt_2025}. For a comparison of derivations for $\bm{b} = 0$ and without the constraint on the enclosed volume we refer to \cite{reuther2015interplay,reuther2018erratum,brandner2022derivations}. 

Considering the overdamped limit, formally letting $\gamma \to \infty$, leads to the classical dynamic equations
\begin{align}
   \label{eq:ol_1}
    \mathbf{u} &= - \nabla_\S p - p \H\bm{\nu} + \bm{b} - \lambda \bm{\nu} \\
   \label{eq:ol_2}
       \div_P \bm{u} &= 0 \\
   \label{eq:ol_3}
   \int_\S \bm{u}\cdot\bm{\nu}\,d\S &= 0 \,,
\end{align}
for an inextensible membrane with constant volume. Using \eqref{eq:ol_1} in \eqref{eq:ol_2} provides the equation for the Lagrange multiplier for the inextensibility constraint $- \Delta_\S p + p \H^2 + \lambda = \bm{b} \cdot \bm{\nu} \H$ and correspond to previous models, if $\bm{b}$ only contains second order geometric terms \cite{biben2003tumbling,KIM20104840,Salac_Miksis_2012,HMLLRV_IJBB_2013,AEV_JCP_2014}.

In contrast with previous approaches for fluid deformable surfaces \cite{Torres-Sanchez_2019,Reuther_2020,Krause_2023,Bachini_2023,Krause_2024} the bending force $\bm{b}$ also contains higher order geometric terms.

\subsection{Bending forces}
The bending force eq. \eqref{eq:bending_force} results from the bending energy 
\begin{equation}\label{eq:bending_energy}
    \energy_{BE}(\meanCurv,  \K ) = \int_\mathcal{\SurfDomain} \densityBE(\meanCurv,  \K )\,d\SurfDomain
\end{equation}
with bending energy density $\densityBE$. Formally we can derive $\densityBE$ via Taylor expansion at the spontaneous curvature $\H_0$ leading to 
\begin{align*}
	\densityBE(\meanCurv,\K)
		=& \sum_{n=0}^{N} \sum_{\alpha=0}^{\lfloor\frac{n}{2}\rfloor} k_{n,\alpha}(\meanCurv-\meanCurv_{0})^{n-2\alpha} \K^\alpha\\
		=& \underbrace{k_{0,0}}_{n=0}
			+ \underbrace{k_{1,0}(\meanCurv-\meanCurv_{0})}_{n=1}
			+ \underbrace{k_{2,0}(\meanCurv-\meanCurv_{0})^2 + k_{2,1}\K}_{n=2}\\
        &+ \underbrace{k_{3,0}(\meanCurv-\meanCurv_{0})^3 + k_{3,1}(\meanCurv-\meanCurv_{0})\K}_{n=3}\\
		&+ \underbrace{k_{4,0}(\meanCurv-\meanCurv_{0})^4 + k_{4,1}(\meanCurv-\meanCurv_{0})^2\K+ k_{4,2}\K^2 }_{n=4}\\
		&+ \ldots
\end{align*}
in terms of geometric orders $ n \le N\in\N $ for different bending rigidity parameters $ k_{n,\alpha}\in\R $. This expression corresponds to the generalized form of the classical Canham/Helfrich energy ($n=2$) introduced in \cite{Mitov_1978} and also considered in \cite{Fournier_1997,Capovilla_2003,Deserno_2015} if gradient terms are neglected. It can be simplified assuming certain properties of the bending energy: As we are interested in variations of the energy, we can omit the constant contribution $k_{0,0}$. We will further not allow for topological changes, thus, we can omit the $k_{2,1}\K $ term, applying Gauss--Bonnet's theorem. Furthermore, we only allow for terms which guarantee boundedness from below. This excludes all odd geometric orders as well as $k_{4,1}$.  Considering these points and only contributions up to geometric order $N=4$ leads to the following bending energy density
\begin{align*}
	\densityBE(\meanCurv,\K)
		=& k_{2,0}(\meanCurv-\meanCurv_{0})^2 + k_{4,0}(\meanCurv-\meanCurv_{0})^4
            + k_{4,2}\K^2,
\end{align*}
which has been considered in \cite{Shemesh_2003} to study the stability of tubular shapes considering $k_{2,0} > 0$, $k_{4,0} = 0$ and $k_{4,2} \gtrless 0$ as parameters to stabilize cylindrical shapes ($k_{4,2} > 0$) or to enforce pearling ($k_{4,2} < 0$). The same form is also considered in \cite{Fournier_1997}, arguing that a cylinder with radius $R$ is stable if $k_{2,0} < 0$ and $k_{4,0} = - \frac{R^2}{2} k_{20}$ with $k_{4,2}$ not determined. We here restrict the parameter space to $k_{2,0} > 0$, $k_{4,0} = 0$ and $k_{4,2} > 0$. We would like to remark that more general parameter combinations are possible, still leading to well-posed bending energies and stable solutions. However, $\K = 0$ is a property of a tube and thus $k_{4,2} \K^2$ seems to be the most plausible higher order extension in the bending energy. In any case the bending force $\bm{b}$ is derived as the negative of the variational derivative of the bending energy \eqref{eq:bending_energy}
\begin{equation*}
    \bm{b} = -\frac{\delta\energy_{BE}}{\delta \param}.
\end{equation*}
A detailed derivation of the bending force is done in Appendix \ref{app:derivation}.

\subsection{Numerical approach}

The numerical approach extends the approach used in \cite{Krause_2023,Bachini_2023}, which is based on surface finite elements \cite{Dziuk_2013,Nestler_2019} and builds on a Taylor-Hood element for the surface Navier-Stokes equations, higher order surface
parametrizations, appropriate approximations of the geometric quantities, mesh redistribution, a semi-implicit discretization in time and an iterative approach to deal with the non-local constraint on the enclosed volume. Additional challenges emerge from the higher order geometric terms. In Appendix \ref{app:numerics} we provide a detailed description and in Appendix \ref{app:validation} convergence studies for these terms. The implementation is done in DUNE/AMDiS \cite{vey2007amdis,witkowski2015software}. Throughout the paper we consider $\Reyn = 1.0$ and $\gamma = 0$ and only vary $\H_0$, $k_{2,0}$ and $k_{4,2}$. Numerical parameters, such as mesh size $h$ and time step $\tau$ are chosen to resolve the highest curvature values and to meet stability constraints, following \cite{Krause_2023}.

\section{Results}\label{s:results}

\subsection{Axialsymmetric simulation without surface viscosity}
\begin{figure}
\begin{tabular}{ l}
    (a) \\
    \includegraphics[width=0.95\columnwidth]{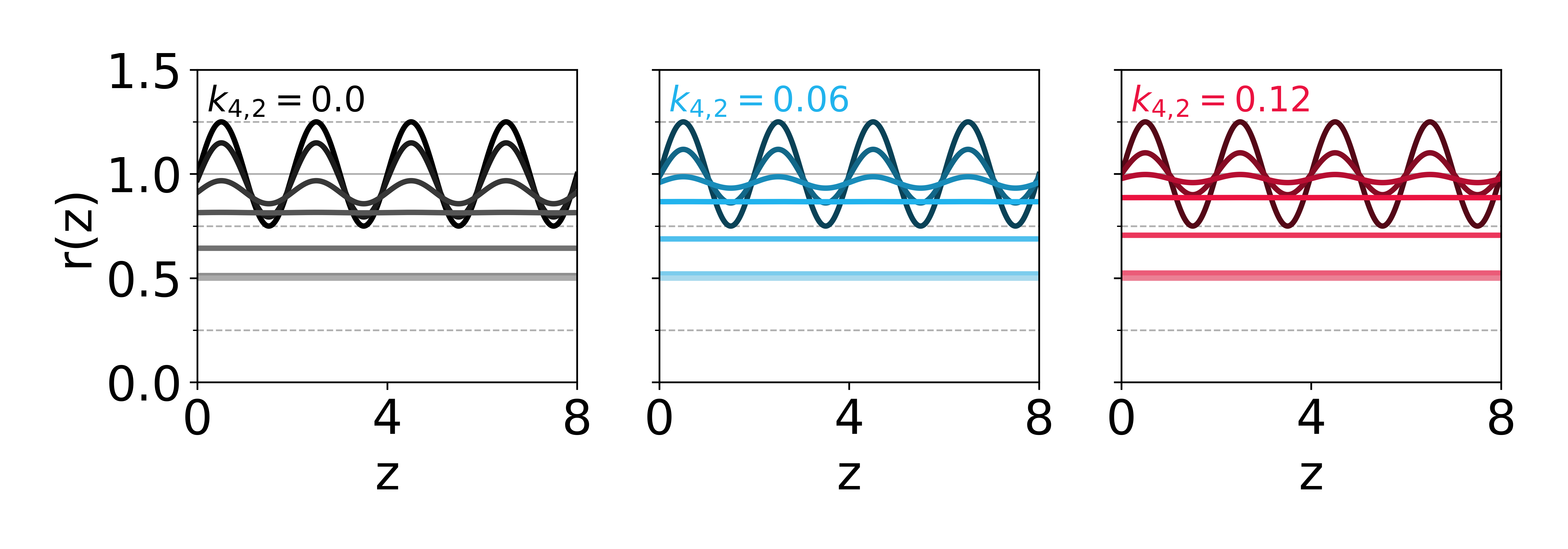}\\
    (b) \\
    \includegraphics[width=0.95\columnwidth]{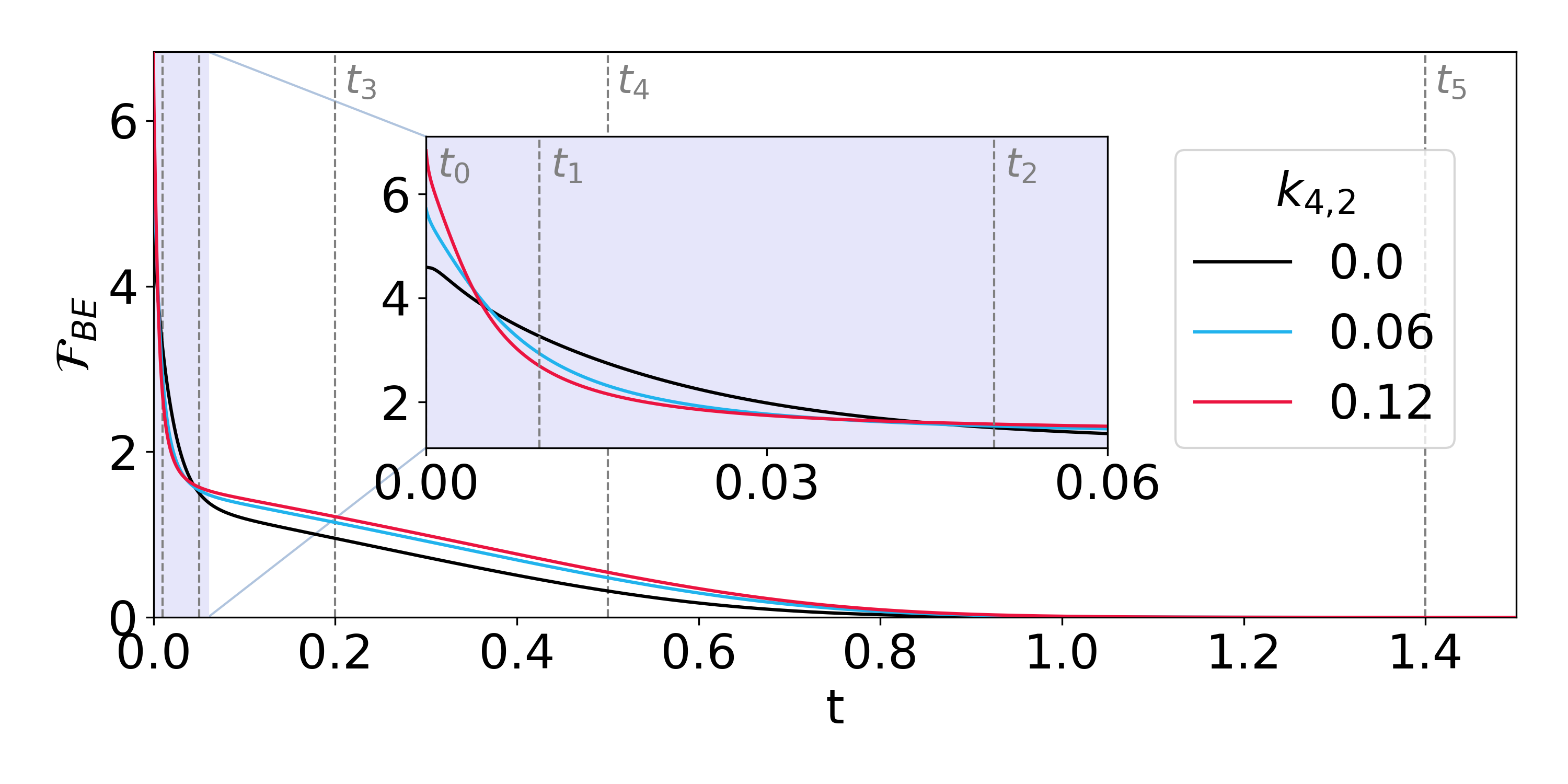}

\end{tabular}
    \caption{(a) Time evolution of axialsymmetric simulations for different parameters for $k_{4,2}$ starting from the same periodic solution $r_0(z) = \tfrac{1}{4}\sin(\pi z)+1$ at $t_0$. The parameters are $k_{4,2} = 0$, $0.06$ and $0.12$. The depicted time instances $t_0=0$, $t_1=0.01$, $t_2=0.05$, $t_3=0.2$, $t_4=0.5$ and $t_5=1.4$ are marked in (b). (b) Time evolution of the bending energy for different values of $k_{4,2}$. Other parameters are $k_{2,0} = 0.2$ and $H_0 = -2$.}
    \label{fig:axi}
\end{figure}

We begin by describing the membrane using cylindrical coordinates $(r,\theta,z)$ and consider a rotational symmetric tube with periodic boundary conditions. With
\begin{align} 
    \H &= \frac{1}{\sqrt{1 + (\partial_z r)^2}}(\frac{\partial_{zz} r}{1 + (\partial_z r)^2} - \frac{1}{r})\,, \\ \K &= - \frac{\partial_{zz} r}{r (1 + (\partial_z r)^2)^2}
\end{align}
the bending energy density $\densityBE(\meanCurv,\K)$ can be reformulated and considered as a function of $r(z,t)$, the radial distance of the axisymmetric membrane from the cylindrical symmetry axis, where $z$ measures the coordinate along that axis and $t$ is time. For the evolution we consider the corresponding equations to eq. \eqref{eq:ol_1} - \eqref{eq:ol_3} but drop the constraints on inextensibility and volume. The resulting equation to solve reads 
\begin{align}
\frac{1}{\sqrt{1 + (\partial_z r)^2}}\partial_t r = u_N = - \frac{\delta\energy_{BE}}{\delta \param} \cdot \bm{\nu} = b\, ,
\end{align}
with $b = \bm{b} \cdot \bm{\nu}$. While lengthy if fully written down, the resulting model can easily be solved. We again use finite elements in space and a semi-implicit discretization in time and consider DUNE/AMDiS \cite{vey2007amdis,witkowski2015software} for the realization. Fig. \ref{fig:axi} shows the time evolution of a periodically perturbed tube for different values of $k_{4,2}$ together with the evolution of the bending energy $\energy_{BE}$ over time.
For $k_{4,2} = 0$ this is a well studied problem of the stability of a tube \cite{RevModPhys.82.1607,refId0,PhysRevLett.120.138102}. The parameters are chosen to remain within the stability region. The results clearly indicate a stronger damping of the perturbations at early times for increasing values of $k_{4,2}$.

Results of a linear stability analysis \cite{PhysRevE.109.044403} for the case of $k_{4,2} = 0$ but considering surface viscosity using a Stokes approximation of eq. \eqref{eq:dyn_sys1} indicate a similar stability region as in the overdamped limit and we assume this to hold also for $k_{4,2} > 0$ and the full problem. 

\subsection{Equilibrium shape of tubular cell}
\begin{figure}
\begin{tabular}{ l l }
    (a) & (b)\\
    \includegraphics[width=0.45\columnwidth]{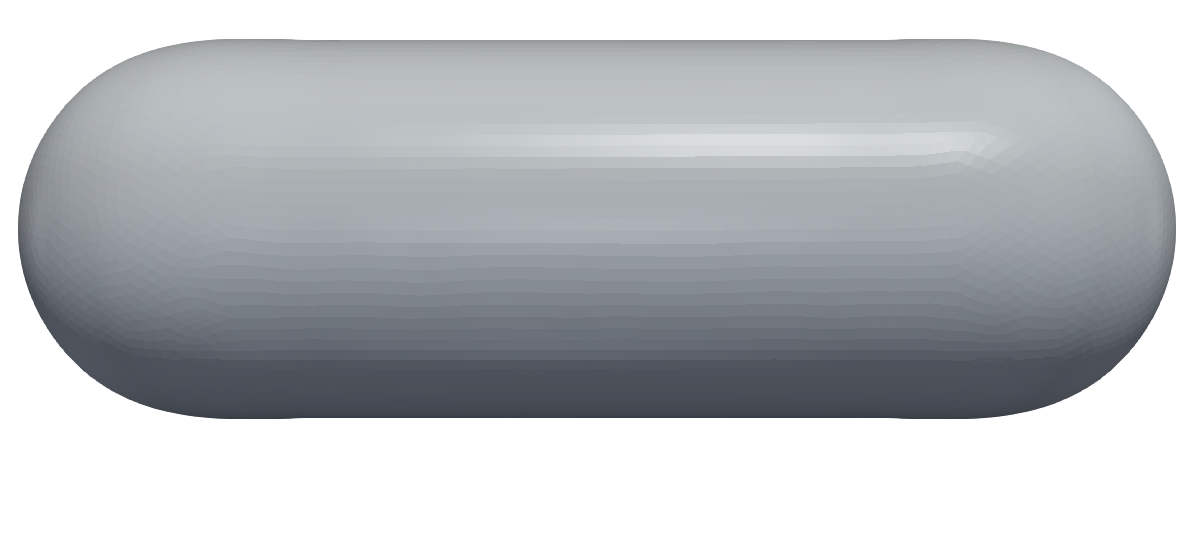}&
    \includegraphics[width=0.45\columnwidth]{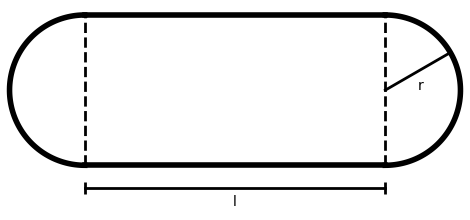}
\end{tabular}
    \caption{(a) As initial surface we consider a cylinder with hemispherical caps. (b) The geometry is determined by the length of the cylinder $l$ and the radius $r$ of the cylinder and hemispherical caps.}
    \label{fig:init_tub}
\end{figure}

The shape of a tubular cell can be approximated by a cylinder with hemispherical caps, as shown in Fig. \ref{fig:init_tub}. For this simple shape, we can compute the bending energy as
\begin{align*}
    \energy_{BE} &= k_{2,0}(2\pi rl(\frac{1}{r}-\meanCurv_0)^2 + 4\pi r^2(\frac{2}{r}-\meanCurv_0)^2) +  k_{4,2} \frac{4\pi}{r^2}
\end{align*}
where $r$ is the radius and $l$ the length of the cylinder. For $\H_0 = \frac{1}{r}$, the energy of the cylindrical part vanishes and the energy simplifies to 
\begin{align*}
    \energy_{BE} = 4\pi(k_{2,0} + \frac{k_{4,2}}{r^2}).
\end{align*}
Note that this energy is independent of the length of the cylinder.
We consider this shape as the initial configuration and the energy as a quantity for comparison.

\begin{figure*}
\begin{tabular}{l}
    (a)\\ 
    \includegraphics[width=0.95\textwidth]{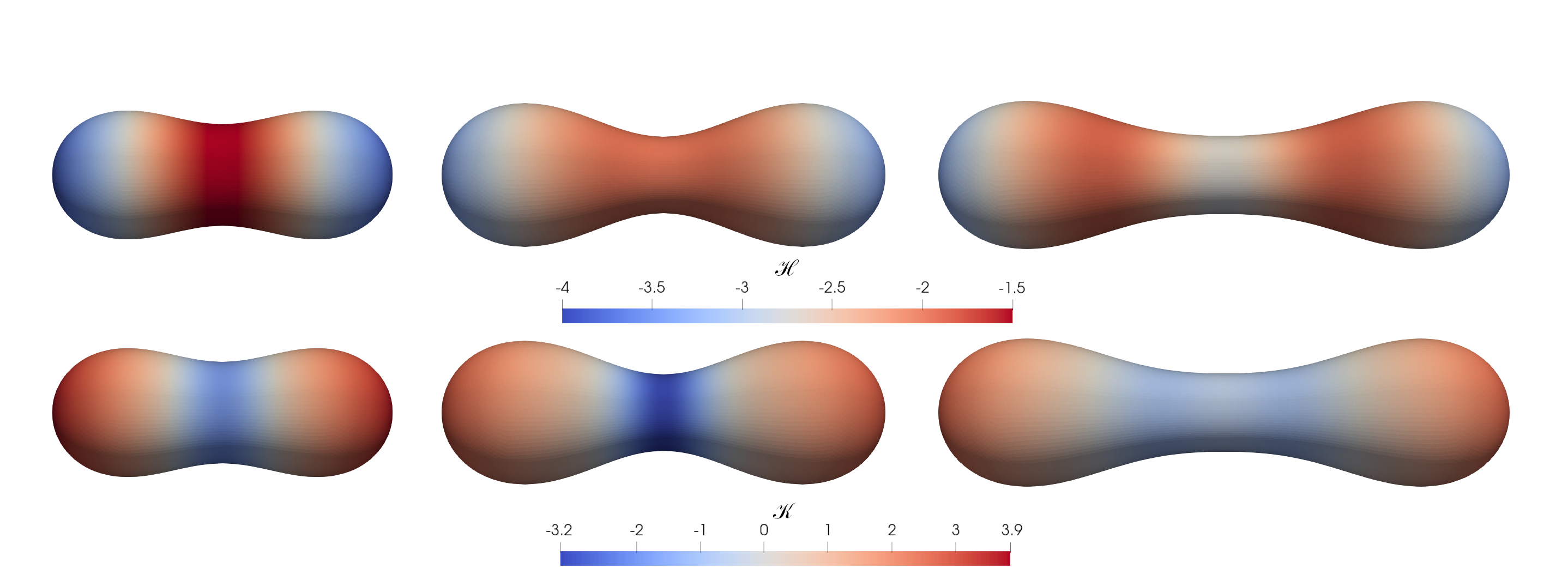}\\
\end{tabular}
\begin{tabular}{l l l}
    (b) \\ 
    \includegraphics[width=0.95\textwidth]{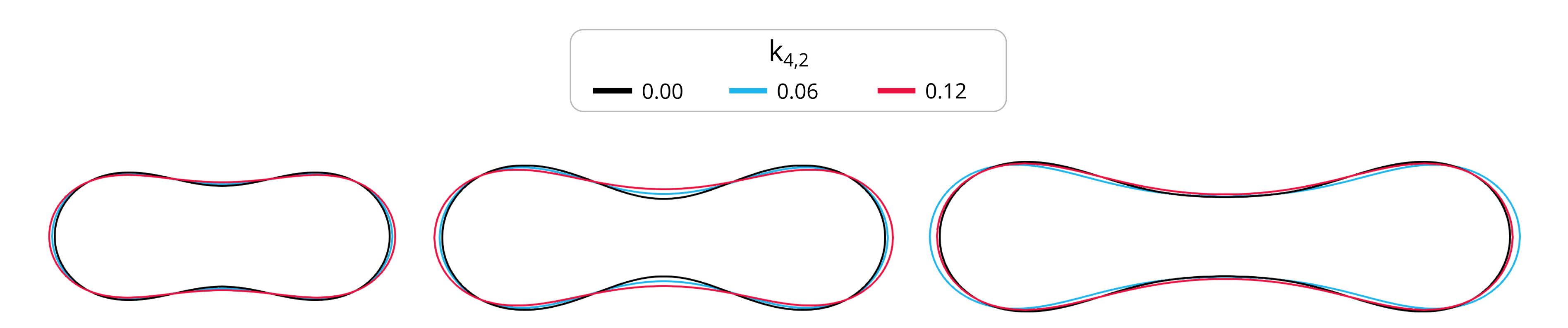}\\
\end{tabular}
    \caption{(a) Equilibrium shapes for $k_{2,0} = 0.125$, $H_0 = -2$ and $k_{4,2} = 0$. As initial condition we consider a cylinder of radius $r = 0.5$ with hemispherical caps and vary the length $l$ of the cylindrical part. From left to right the lengths are $l = 2$, $3$ and $4$. The surfaces are colored by mean curvature $\H$ (upper row) and Gaussian curvature $\K$ (lower row). (b) Equatorial cuts through the equilibrium shapes for different values of $k_{4,2}$ and $l$, keeping $k_{2,0} = 0.125$ and $H_0 = -2$. }
    \label{fig:final_tub}
\end{figure*}

\begin{figure}
\begin{tabular}{l  l}
    (a) & (b)  \\
    \includegraphics[width=0.45\columnwidth]{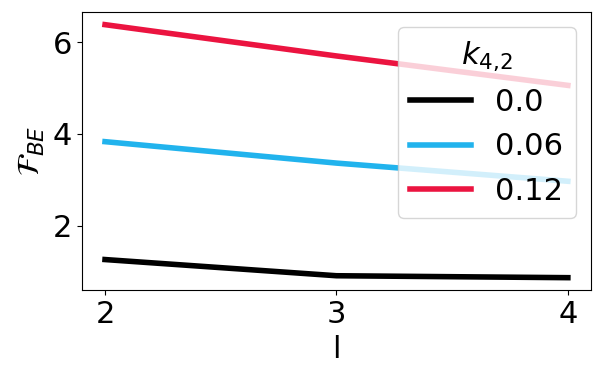}&
    \includegraphics[width=0.45\columnwidth]{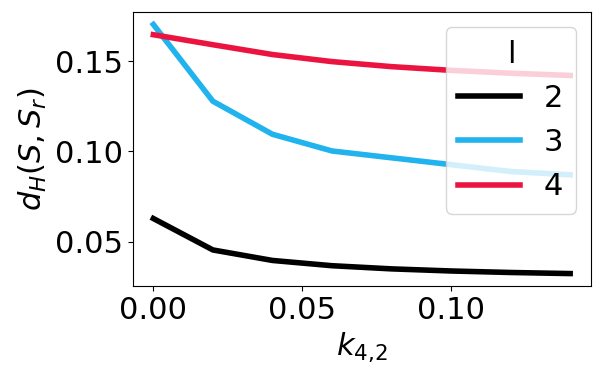}
\end{tabular}
    \caption{(a) Bending energy of the final shape for fixed values of $k_{42}$ for different length $l$ (b) Hausdorff distance between the equilibrium shape and the initial shape for different lengths $l = 2$, $3$ and $4$ as a function of $k_{4,2}$. }
    \label{fig:plots_tub}
\end{figure}

Fig. \ref{fig:final_tub} shows the final configurations achieved by solving the full problem, eq. \eqref{eq:dyn_sys1} - \eqref{eq:dyn_sys3}, for three different values $l$. In all cases, the shape deviates from the initial configuration. For $k_{4,2} = 0$ the resulting shapes correspond to the equilibrium prolate shapes in \cite{Seifert1991}. However, for $k_{4,2} > 0$, these shapes deviate. The bending energy is reduced by increasing $l$ and the Hausdorff distance $d_H(\S,\S_0)$ to the idealized initial shape of a cylinder with hemispherical caps $\S_0$ is reduced by increasing $k_{4,2}$, see Fig. \ref{fig:plots_tub}. These trends are far from being general or monotonous. This can also not be expected due to the highly nonlinear coupling of the geometric terms and the considered constraints on volume and inextensibility. However, they confirm the intuition of the potential impact of $k_{4,2} \K^2$ in the bending energy on the emerging equilibrium shapes. Within the stability region, higher order geometric terms related to $k_{4,2} \K^2$ seem to stabilize tubular shapes.

\subsection{Dynamic evolution}

\begin{figure*}
    \begin{tabular}{l}
    (a)\\ \includegraphics[width=0.95\textwidth]{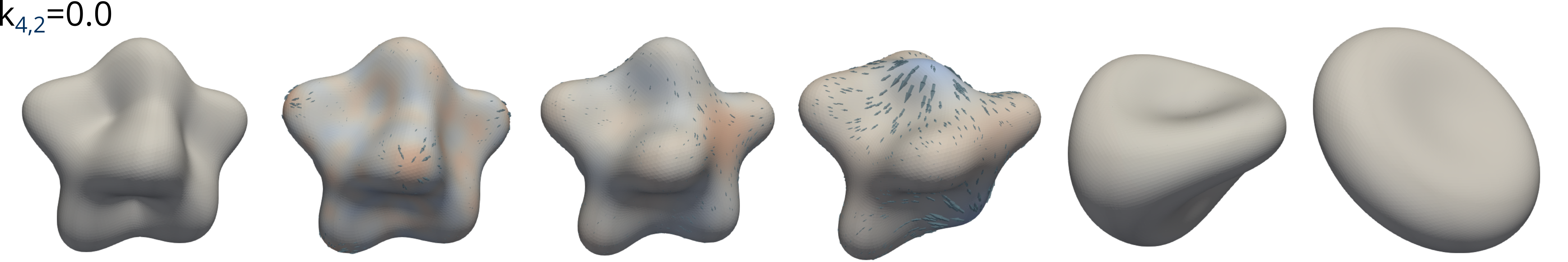}\\
    (b)\\ \includegraphics[width=0.95\textwidth]{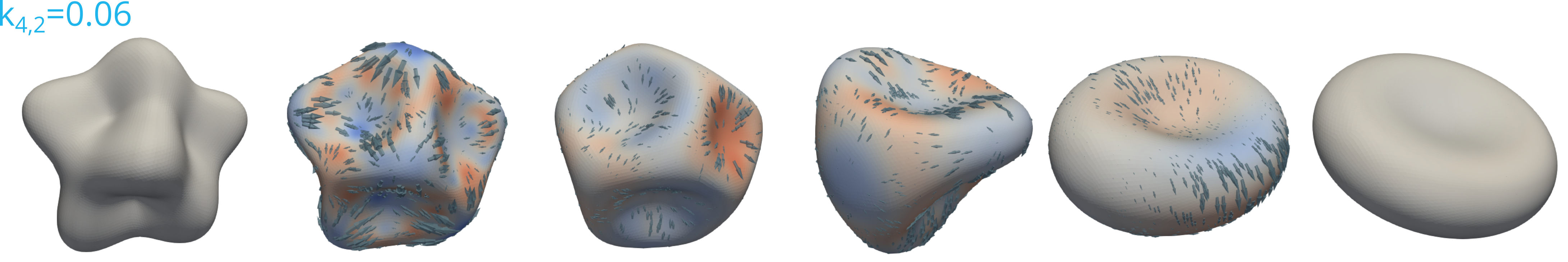}\\
    (c)\\ \includegraphics[width=0.95\textwidth]{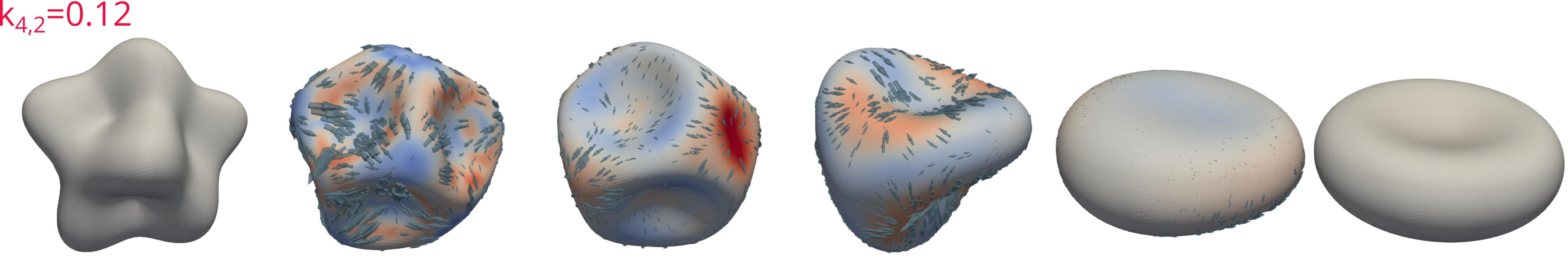}\\
    (d)\\ 
    \includegraphics[width=0.95\textwidth]{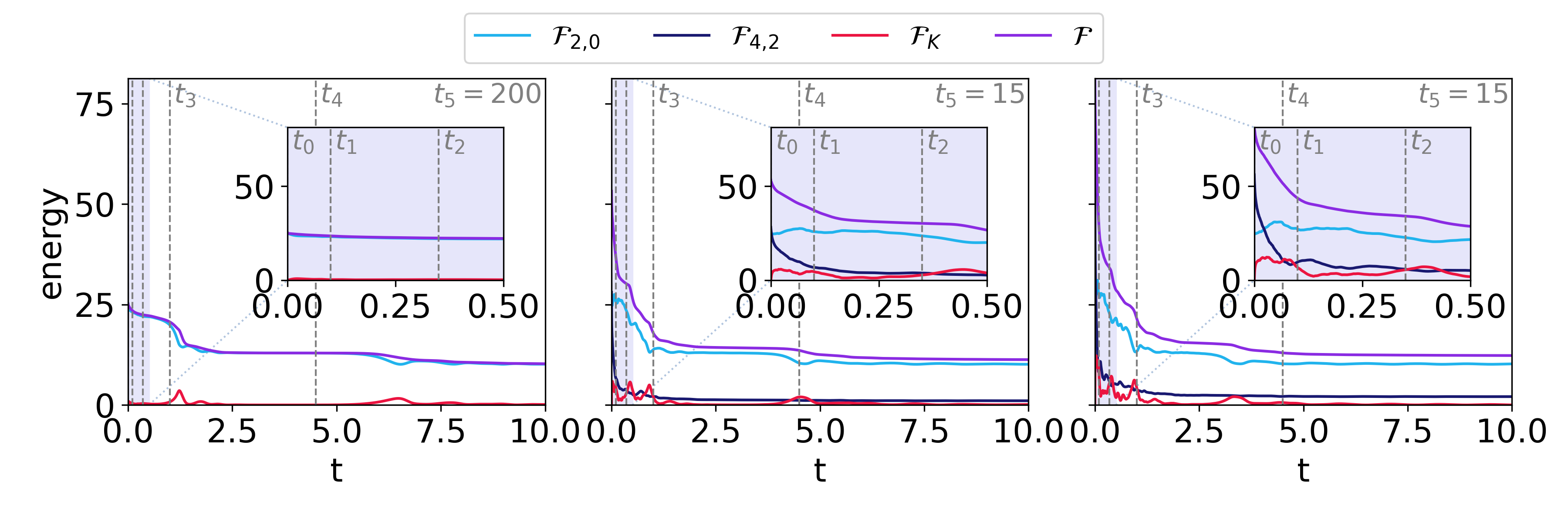} 
    \end{tabular}

    \caption{Evolution of a perturbed sphere considering parameters $k_{2,0} = 0.125$, $H_0 = 0$ and different values of $k_{4,2}$. (a)-(c) Time instances for the evolution for $k_{4,2} = 0$, $0.06$ and $0.12$, respectively. The color coding indicates movement in normal direction (red - movement outwards, blue - movement inwards) and the arrows indicate the tangential velocity. The time evolves from left to right, the time instances except for the last one are equal and are depicted in (d). (d) Evolution of the different energy contributions, $\mathcal{F}_{2,0}$ and $\mathcal{F}_{4,2}$ are the energies linked to the corresponding bending terms, $\mathcal{F}_K$ is the kinetic energy. The bending energy is $\mathcal{F}_\text{BE} = \mathcal{F}_{2,0} + \mathcal{F}_{4,2}$ and the total energy $\mathcal{F}= \mathcal{F}_\text{BE} + \mathcal{F}_K$. Shown is the evolution for $k_{4,2} = 0$, $0.06$ and $0.12$ (from left to right). Within \dag corresponding videos to the evolution in (a) - (c) are provided using a LIC filter for visualization of the tangential flow.}
    \label{fig:dynamics}
\end{figure*}

In order to further explore the influence of the higher order geometric term related to $k_{4,2} \K^2$ on the dynamics, we consider an initial surface $\SurfDomain_0$ as a perturbed unit sphere
\begin{align*}
\SurfDomain_0 = \left\{1+r_0Y^m_l(\phi,\vartheta): \phi \in [0,\pi],\vartheta\in[-\pi,\pi] \right\}, \quad r_0>0, \\
Y^m_l(\phi,\vartheta) = \sqrt{\frac{2l+1(l-m)!}{4\pi(l+m)!}}P^m_l(\cos\vartheta)e^{im\phi}
\end{align*}
with spherical harmonics $Y^m_l$ and Legendre polynomials $P^m_l$ and the case $l = 5$, $m = 3$ and $r_0 = 0.5$. The velocity field is initialized with $\vectvel_0 = 0$. The initial surface has non-zero mean and Gaussian curvature and is out of equilibrium. The resulting bending force induces shape deformations in normal direction. However, the curvature terms also induce tangential flows, which also contribute to shape deformations. This coupling between tangential flow and shape deformations is well understood and shown to enhance the evolution towards equilibrium shapes \cite{Krause_2023}. Here, we explore the evolution for different values of $k_{4,2}$, which read $k_{4,2} = 0$, $0.06$ and $0.12$.
Fig. \ref{fig:dynamics} shows snapshots of the evolutions. The color coding corresponds to shape deformations (red - movement outwards, blue - movement inwards) and the arrows indicate the tangential velocity, where the length scales with the magnitude. Furthermore, the energy contributions are shown over time. Here, the bending energy is split into the different contributions $\energy_{BE} = \energy_{2,0} + \energy_{4,2}$ and the total energy $\energy = \energy_{BE} + \energy_K$ is the sum of the bending energy and the kinetic energy $\energy_K = \int_\S \frac{1}{2} \vectvel^2 \, d \S$. All evolutions converge to the equilibrium shape, which for the case $k_{4,2}= 0$ again corresponds to the prolate shapes in \cite{Seifert1991}. The other equilibrium shapes only slightly differ. The difference in orientation might result from the different dynamics. However, due to the decoupling from the surrounding bulk phases and the considered parameter $\gamma = 0$, force-free rigid body rotations are also possible \cite{nestler2023stability}. While the equilibrium shapes are similar, significant changes can be observed in the dynamics. The close coupling between the bending energy $\energy_{BE}$ and the kinetic energy $\energy_{K}$ can be observed and related to significant shape changes. But their appearance differs. The plateau in $\energy_{BE} = \energy_{2,0}$ for $k_{4,2} = 0$ between $t \approx 2$ and $t \approx 6$ is reduced for $k_{4,2}= 0.06$ and $0.12$ and already ends at $t \approx 4.5$ and $t \approx 3$, respectively. It should be noted that the absolute values of the energies cannot be directly compared as $k_{4,2}$ varies in the definition. However, qualitatively the higher order geometric terms further enhance the evolution and lead to alternative pathways to dissipate energy. This is most pronounced in the inlets highlighting the initial evolution, the drastic decrease of $\energy_{4,2}$ is associated with large fluctuation of $\energy_K$. This behaviour increases with increasing values of $k_{4,2}$. Furthermore, while $\energy_{4,2}$ is roughly one order of magnitude smaller than $\energy_{2,0}$, the influence on the dynamics is dramatic. 

\section{Conclusions}\label{s:conclusion}

Motivated by rapid shape changes of cells, where an excess of membrane that is organized in membrane reservoirs is made available to the cell in the order of seconds \cite{DeBelly_Cell_2023}, we formulated a mesoscopic membrane model which facilitates this behavior. This, on the one side includes an extension of the classical Canham-Helfrich model towards higher order geometric terms, and on the other side the explicit treatment of the fluid properties of the membrane by considering membranes as fluid deformable surfaces. The first aspect adds to the classical bending energy $\energy_{2,0} = \int_\S k_{2,0} (\H - \H_0)^2 \; d \S$ a term proportional to the Gaussian curvature squared, $\energy_{4,2} = \int_\S k_{4,2} \K^2 \, d \S$. This not only helps to damp perturbations of tubes, it also has a tendency to stabilize them. This has been considered in idealized rotational symmetric and full three-dimensional situations by analyzing the evolution and the emerging equilibrium shapes. If combined with the second aspect, which takes the surface viscosity of the membrane into account and combines the bending in normal direction with the properties of an inextensible surface fluid, as a fluid deformable surface, the dynamic drastically changes. Already for the classical bending energy $\energy_{2,0} = \int_\S k_{2,0} (\H - \H_0)^2 \; d \S$ an enhanced evolution towards the equilibrium shape has been observed if the effects of surface viscosity are taken into account \cite{Krause_2023}. With the higher order geometric terms $\energy_{4,2} = \int_\S k_{4,2} \K^2 \, d \S$ this is further enhanced. The considered numerical experiments for the relaxation of a perturbed sphere showed alternative pathways to dissipate energy and strong tangential flows inducing fast shape changes.
While certainly more research is needed to fully explore the potential of the higher order geometric terms, e.g., with respect to the stability of tubes extending the analysis in\cite{PhysRevE.109.044403}, the numerical studies already clearly indicate the potential for rapid shape changes. Even if only passive contributions of the membrane are considered, and a full model for morphological changes of a cell requires to also take the active processes of the underlying cortex and probably even more phenomena into account, the study contributes to identifying underlying general mechanical principles which might help to predict and control the dynamics of cells\cite{10.1242/jcs.260744}.  

\section*{Acknowledgements}

We acknowledge fruitful discussions with Patrick Zager (UCSF) and Veit Krause (TUD). This work was funded by DFG within FOR3013 "Vector- and tensor-valued surface PDEs". We further acknowledge computing resources at FZ Jülich under Grant No. MORPHO and at ZIH under Grant No. WIR.

\bibliography{lit} 
\bibliographystyle{rsc} 

\section{Appendix}

\subsection{Model derivation}\label{app:derivation}

We here compute the variational derivatives of the higher order geometric terms in the bending energy. We therefore write the bending energy as
\begin{align*}
	\energy_{BE} 
		&= \int_{\S} k_{2,0} \densityBE^{2,0} + k_{4,0} \densityBE^{4,0} + k_{4,2} \densityBE^{4,2}\,d\S
\end{align*}
with bending density components $ \densityBE^{n,\alpha} = (\H-\H_{0})^{n-2\alpha} \K^{\alpha} $.
The corresponding forces $ \bm{b}^{n,\alpha}\in\tangentR $ are given by
\begin{align*}
	\innerLTwo{\tangentR}{\bm{b}^{n,\alpha}, \bm{W}}
		&:= - \innerLTwo{\tangentR}{\frac{\delta}{\delta\param}\int_{\S}\densityBE^{n,\alpha}\,d\S , \bm{W}}
\end{align*}
for all $ \bm{W} \in \tangentR $. The result for $\densityBE^{2,0}$ is well known
\begin{align*}
    \bm{b}^{2,0}&= -2(\Delta_\S\H + (\H-\H_0)(\|\shapeOp\|^2-\frac{1}{2}\meanCurv\left(\meanCurv-\meanCurv_0\right)))\normalvec \formPeriod
\end{align*}
For the remaining terms we use the deformation derivative $ \eth_{\Wb} $ \cite{NitschkeSadikVoigt_A_2022} and obtain the deformation formula
\begin{align}\label{eq:deformationFormula}
	\innerLTwo{\tangentR}{\bm{b}^{n,\alpha}, \Wb}
		= -\int_{\S} \eth_{\Wb} \densityBE^{n,\alpha} + \densityBE^{n,\alpha}\DivC\Wb\, d\S \,.
\end{align}
Moreover, it is
\begin{align}
	\ProjMat(\eth_{\Wb}\shapeOp)\ProjMat \label{eq:deformation_shop}
		&= \GradSurf(\nub\cdot\GradC\Wb) - \shapeOp\GradC\Wb \formComma\\
	\eth_{\Wb}\meanCurv
		&= \Tr(\ProjMat(\eth_{\Wb}\shapeOp)\ProjMat) \label{eq:deformation_meanc}
		 = \divS(\nub\cdot\GradC\Wb) - \shapeOp\dbdot\GradC\Wb
\end{align}
valid\cite{NitschkeSadikVoigt_A_2022}. As a consequence, the deformation formula \eqref{eq:deformationFormula} and integrations by parts result in
\begin{align*}
	\innerLTwo{\tangentR}{\bm{b}^{4,0}, \Wb} \hspace{-5em}\\
		&= - \int_{\S}  4(\meanCurv-\meanCurv_0)^3 \eth_{\Wb} \meanCurv + (\meanCurv-\meanCurv_0)^4\DivC\Wb\, d\S\\
		&= \int_{\S}  12 (\meanCurv-\meanCurv_0)^2 (\GradSurf\meanCurv)\cdot(\nub\cdot\GradC\Wb)\\
		&\quad\quad+(\meanCurv-\meanCurv_0)^3 (4\shapeOp-(\meanCurv-\meanCurv_0)\ProjMat )\dbdot\GradC\Wb \, d\S
\end{align*}
Another application of integrations by parts \wrt\ $ \GradC $, yields
\begin{align*}
	\bm{b}^{4,0}
		&= -\DivC\big( 12(\meanCurv-\meanCurv_0)^2 \nub\otimes\GradSurf\meanCurv \\
		&\hspace{4em} + (\meanCurv-\meanCurv_0)^3 (4\shapeOp-(\meanCurv-\meanCurv_0)\ProjMat ) \big)\formComma
\end{align*}
where
\begin{align}
	\DivC(\bm{\sigma}\ProjMat) \label{eq:DivC_decomp}
		&= \divS(\ProjMat\bm{\sigma}\ProjMat) - \nub\cdot\bm{\sigma}\shapeOp
			+\left( \divS(\nub\cdot\bm{\sigma}\ProjMat) + \shapeOp\dbdot\bm{\sigma}\right)\nub
\end{align}
holds for all $ \bm{\sigma}\in\tangentR[2] $\cite{NitschkeSadikVoigt_A_2022,NitschkeVoigt_2025}.
Therefore, the tangential part of $ \bm{b}^{4,0} $ cancels out and we obtain
\begin{align*}
	\bm{b}^{4,0}
		&= - ( 12\divS((\meanCurv-\meanCurv_0)^2\GradSurf\meanCurv)\\ 
		&\hspace{3em} + (\meanCurv-\meanCurv_0)^3((3\meanCurv+\meanCurv_0)\meanCurv- 8\GaussCurv) ) \nub\formPeriod
\end{align*}
With \eqref{eq:deformation_shop} and \eqref{eq:deformation_meanc}, the deformation derivative of $ \GaussCurv = 2(\meanCurv^2 - \|\shapeOp\|^2) $ reveals
\begin{align*}
	\eth_{\Wb}\GaussCurv
		&= \meanCurv\eth_{\Wb}\meanCurv - \shapeOp\dbdot\eth_{\Wb}\shapeOp\\
		&= (\meanCurv\ProjMat-\shapeOp) \dbdot (\GradSurf(\nub\cdot\GradC\Wb) - \shapeOp\GradC\Wb) \formPeriod
\end{align*}
As a consequence, the deformation formula \eqref{eq:deformationFormula}, integrations by parts, and $ \shapeOp^2 = \meanCurv\shapeOp - \GaussCurv\ProjMat $, result in
\begin{align*}
	\innerLTwo{\tangentR}{\bm{b}^{4,2}, \Wb} \hspace{-6em}\\
		&= - \int_{\S}  2\GaussCurv \eth_{\Wb} \GaussCurv + \GaussCurv^2\DivC\Wb\, d\S\\
		&= \int_{\S} 2 \divS\left( \GaussCurv(\meanCurv\ProjMat-\shapeOp) \right) \cdot (\nub\cdot\GradC\Wb)
						+\GaussCurv^2\ProjMat\dbdot\GradC\Wb\, d\S \formPeriod
\end{align*}
Since $ \divS( \meanCurv\ProjMat-\shapeOp ) = 0 $ holds, integrations by parts yields
\begin{align*}
	\bm{b}^{4,2}
		&= -\DivC\left( 2\nub\otimes(\meanCurv\ProjMat-\shapeOp)\GradSurf\GaussCurv  + \GaussCurv^2\ProjMat \right)\formPeriod
\end{align*}
Using \eqref{eq:DivC_decomp}, the tangential part of $ \bm{b}^{4,2} $ cancels out and we obtain
\begin{align*}
	\bm{b}^{4,2}
		&= -\left( 2\divS((\meanCurv\ProjMat-\shapeOp)\GradSurf\GaussCurv) + \meanCurv\GaussCurv^2 \right)\nub\formPeriod
\end{align*}
Putting everything together yields eq. \eqref{eq:bending_force}. For the derivation of the other parts of the model we refer to \cite{Bachini_2023,NitschkeVoigt_2025}.

\subsection{Numerical method}\label{app:numerics}
We consider a surface finite element method (SFEM) \citep{Dziuk_2013,Nestler_2019} to solve the highly nonlinear set of geometric and surface partial differential equations \eqref{eq:dyn_sys1}-\eqref{eq:dyn_sys3}, using the approaches in \cite{Krause_2023,Bachini_2023}. 

We combine the system \eqref{eq:dyn_sys1}-\eqref{eq:dyn_sys3} with a mesh redistribution approach, see \citet{Barrett_SIAMJSC_2008}. These are equations for the parametrization
\begin{align}
        \DerT \param \cdot \normalvec &= \vectvel\cdot\normalvec \label{eq:normalvel}\\
        \meanCurv \normalvec &= \Delta_C \param \,, \label{eq:meandiff}
\end{align}
which generate a tangential mesh movement to maintain the shape regularity and additionally provide an implicit representation of the mean curvature $\meanCurv$. We consider a discrete $k$-th order approximation $\SurfDomain_h^k$ of $\SurfDomain$, with $h$ the size of the mesh elements, i.e. the longest edge of the mesh. We use the DUNECurvedGrid library \citet{praetorius2022dune} and consider each geometrical quantity like the normal vector $\normalvec_h$, the shape operator $\shapeOp_h$, the Gaussian curvature $\K_h$, and the inner products $(\cdot , \cdot)_h$ with respect the $\SurfDomain_h^k$. In the following, we will drop the index $k$. We define the discrete function spaces for scalar functions by $V_{k}(\SurfDomain_h)=\{ \psi \in C^0(\SurfDomain_h) \vert \psi\vert_{\Cell}\in\mathcal{P}_{k}(\Cell)\}$ and for vector fields by $\boldsymbol{V}_{k}(\SurfDomain[\meshparam])=[V_{k}(\SurfDomain[\meshparam])]^3$. Within these definitions $T$ is the mesh element and $\mathcal{P}_{k}$ are the polynomials of order $k$. We consider  $\vectvel_h,\param\in\boldsymbol{V}_3(\SurfDomain_h)$,  $\meanCurv_h \in V_3(\SurfDomain_h)$, and $p_h\in V_2(\SurfDomain_h)$, which leads to an isogeometric setting for the velocity and a $\mathcal{P}_{3}-\mathcal{P}_{2}$ Taylor-Hood element for the surface Navier-Stokes-like equations. We discretize in time using constant time stepping with step size $\tau$. In each time step we solve the surface Navier-Stokes-like equations and the mesh redistribution together. We define a discrete surface update variable $\update^{n}=\param^{n}-\param^{n-1}$, which is considered as unknown instead of the surface parametrization $\param^{n}$. The system to solve reads: 

\newpage
\begin{widetext}
Find
$(\vectvelApprox^n,\pressApprox^n,\meanCurv_h^n,\update^n)\in[\boldsymbol{V}_3\times V_2\times V_3 \times \boldsymbol{V}_3](\SurfDomain_h^{n-1})$ such that:
\begin{align*}
  \frac{1}{\tau} \InnerApprox{\vectvelApprox^{n}-\vectvelApprox^{n-1}}{\TestVecApprox} + \InnerApprox{\GradSurfConv{\vectvelV_{\meshparam}^{n-1}}\vectvelApprox^n}{\TestVecApprox} =& \InnerApprox{\pressApprox^n}{\DivP\TestVecApprox}
   - \frac{2}{\Reyn} \InnerApprox{
   \StressTens(\vectvelApprox^{n})}{\GradP\TestVecApprox} -\gamma \InnerApprox{\vectvel_h^n}{\TestVecApprox} \\
   & + \InnerApprox{ \left(2 k_{2,0}\GradSurf\meanCurv_h^n\right) }{\GradSurf(\TestVecApprox\cdot\normalvec_h^{n-1})} \\
   & + \InnerApprox{2 k_{2,0}(\meanCurv_h^n-\meanCurv_0^{n-1})B^{n-1} \normalvec_h^{n-1}}{\TestVecApprox} + \lambda \InnerApprox{\normalvec_h^{n-1}}{\TestVecApprox} \\
   & + \InnerApprox{k_{4,0}(\H_h^n-\H_0^{n-1})(\H_h^{n-1}-\H_0^{n-1})^2\left(4\shapeOp^{n-1}-(\H_h^{n-1}-\H_0^{n-1})\ProjMat\right)}{\GradC\TestVecApprox}\\
   &+\InnerApprox{12k_{4,0}(\H_h^{n-1}-\H_0^{n-1})^2\normalvec_h^{n-1}\GradSurf\meanCurv}{\GradC\TestVecApprox}\\
   & + \InnerApprox{2k_{4,2}\H_h^n\normalvec_h^{n-1}\otimes(\ProjMat\GradSurf\K^{n-1}_h)}{\GradC\TestVecApprox} \\
   & -\InnerApprox{2k_{4,2}\normalvec^{n-1}_h\otimes (\shapeOp^{n-1}\GradSurf\K^{n-1}_h)}{\GradC\TestVecApprox} + \InnerApprox{k_{42}(\K_h^{n-1})^2\ProjMat}{\GradC\TestVecApprox}\\
\InnerApprox{\DivP\vectvelApprox^{n}}{\TestPressApprox} =& \; 0 \\
    \frac{1}{\tau}\InnerApprox{\update^n \cdot \normalvecApprox^{n-1}}{\TestHApprox} =& \InnerApprox{\vectvel_h^n\cdot\normalvecApprox^{n-1}}{\TestHApprox} \\
    \InnerApprox{\meanCurv_h^n\normalvec_h^{n-1}}{\TestZApprox} 
    + \InnerApprox{\GradC \update^n}{\GradC \TestZApprox} =&
    - \InnerApprox{\GradC \MapU^{n-1}}{\GradC \TestZApprox}
\end{align*}
for all $(\TestVecApprox,\TestPressApprox,\TestHApprox,\TestZApprox)\in[\boldsymbol{V}_3\times V_2\times V_3\times\boldsymbol{V}_3](\SurfDomain_h^{n-1})$, where
$B^{n-1}=\left(\Vert\shapeOp_h^{n-1}\Vert^2-\frac{1}{2}\tr \shapeOp_h^{n-1}(\tr\shapeOp_h^{n-1}-\meanCurv_0(\phi_h^n)) \right)$.
\end{widetext}

In the above formulation we used the identity $\InnerApprox{-\GradSurf p_h^n - p_h^n \meanCurv_h^n \normalvec_h}{\TestVecApprox} = \InnerApprox{p_h^n}{\DivP \TestVecApprox}$. Note that the Lagrange multiplier $\lambda$ is unknown, which leads to an underdetermined system. To resolve that problem, we follow the approach introduced in \cite{Krause_2023}.  In order to fulfill the volume constraint, $\lambda$ has to be chosen such that 
\begin{equation}
	\Phi(\lambda) := \int_{\S^{n-1}_h} \mathbf{u}^{n}_h(\lambda)\cdot \bm{\nu}_h\,d\S = 0.
\end{equation}
We consider $\Phi(\lambda) = 0$ as an equation in $\lambda$ and apply a Newton iteration $\lambda^{j+1} = \lambda^{j} - \Phi(\lambda^{j}) / \Phi'(\lambda^{j})$. After convergence is achieved the new surface $\S_h^n$ needs to be computed by updating the paramerization $\param^{n} =  \param^{n-1} + \update^{n}$, lifting the solutions $\vectvel_h^n$, $\press_h^n$ and $\meanCurv_h^n$ to the new surface and computing the remaining geometric quantities $\normalvec_h^{n}$, $\shapeOp_h^n$, $\nabla_\S \H_h^n$, $\K_h^n$ and $\nabla_\S \K_h^n$ for the new surface. While this approach showed the (expected) optimal order of convergence \cite{Krause_2023}, with respect to computational and numerical analysis results for the underlying surface Stokes equations on stationary surfaces \cite{brandner2022finite,hardering2023tangential} for bending terms up to second order, the approach is not sufficient if higher order geometric terms are included. 

We therefore introduce a smoothing step of the surface quantities $\meanCurv_h^n$, $\nabla_\S\meanCurv_h^n$, $\K_h^n$ and $\nabla_\S\K_h^n$. For each surface quantity or its components $a_h = \meanCurv_h, [\nabla_\S\meanCurv_h^n]_i, \K_h$ and $ [\nabla_\S\K_h]_i$, $i=1,2$ we solve one time step of the diffusion equation
\begin{align*}
    a_s - \epsilon \Delta_\S a_s = a_h,  
\end{align*}
where $\epsilon>0$ is a smoothing parameter and $a_s$ the smoothed surface quantity.

\subsection{Validation}\label{app:validation}
\begin{figure}
\begin{tabular}{ l }
    (a) \\
    \includegraphics[width=0.95\columnwidth]{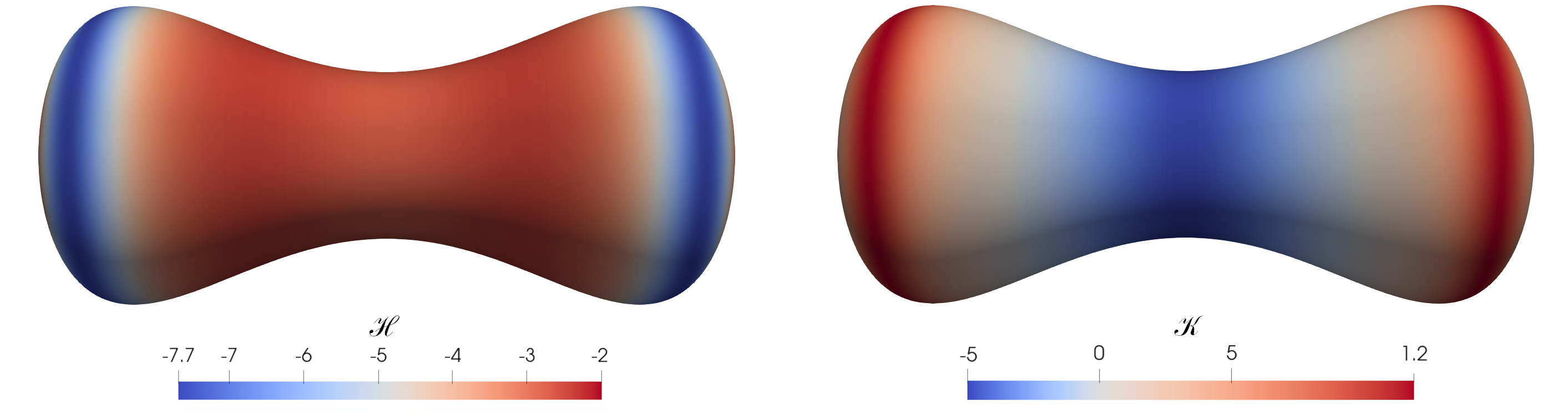}
\end{tabular}
\begin{tabular}{ l }
    (b) \\ \includegraphics[width=0.9\columnwidth]{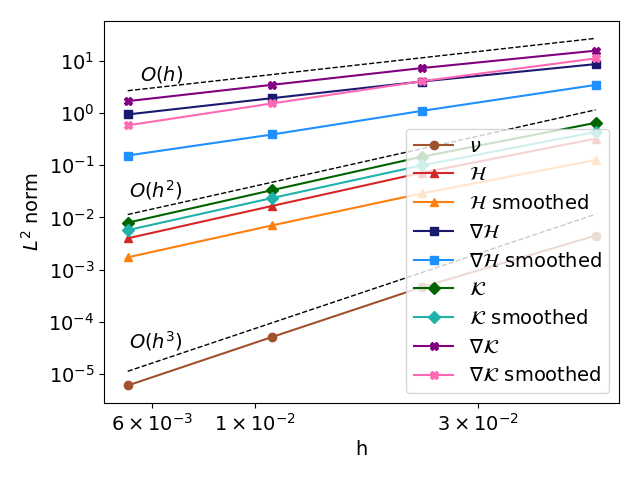}
\end{tabular}
    \caption{(a) Reference surface for the tests. Depicted are the mean curvature $\meanCurv$ and Gaussian curvature $\K$. (b) Errors of geometric quantities with and without additional smoothing step for different grid widths $h$. The orders of convergence are indicated by the dashed lines and are optimal orders for the numerical implementation.}
    \label{fig:convergence}
\end{figure}

Instead of a full convergence study of the numerical approach we only test the smoothing of surface quantities. We consider a surface for which the surface quantities can be computed analytically. The surface is parametrized by $\mathbf{X}:[0,2\pi)\times [0,\pi)\to\R^2$,
\begin{align*}
    \mathbf{X}(\phi, \theta) =
    \begin{pmatrix}
        \frac{1}{4} + \frac{3}{4}\cos^2\theta\sin\phi\sin\theta\\ 
        \cos\phi\left(\frac{1}{4}+\frac{3}{4}\cos^2\theta\sin\theta\right)\\
        \cos\theta
    \end{pmatrix} \, .
\end{align*}
We compute the $L^2$-error of the surface quantities $\meanCurv, \K, \nabla_{\SurfDomain}\meanCurv$ and $\nabla_{\SurfDomain} \K$ for different grid width $h$. The smoothing parameter $\epsilon$ is chosen experimentally such that it shows optimal results. Fig. \ref{fig:convergence} shows that for this test case the surface quantities converge with the optimal orders and that the additional smoothing step for all quantities improves the approximation.
Together with the convergence studies in \cite{Krause_2023,Bachini_2023} these results provide enough confidence in the numerical approach for the full problem including the higher order geometric terms. They require an appropriate resolution of $\K$ and $\nabla_\S \K$, which is achieved in $O(h^2)$ and $O(h)$, respectively. This motivated the considered discrete function spaces.
\end{document}